\documentclass[aps,prd,nofootinbib,superscriptaddress,twoside,twocolumn,floatfix,a4paper,subcaption,reprint,preprintnumbers,showkeys]{revtex4-2}

\usepackage{orcidlink,amssymb,amsmath,hyperref,siunitx,makecell}
\usepackage{enumitem}
\usepackage{ytableau,graphicx,youngtab}
\usepackage{bm}
\usepackage{multirow}
\allowdisplaybreaks

\hypersetup{pdfnewwindow=true,
colorlinks=true,linkcolor=blue,
citecolor=blue,filecolor=blue,urlcolor=blue
}

\newcommand{\ba}{\begin{eqnarray}}
\newcommand{\ea}{\end{eqnarray}}
\newcommand{\bea}{\begin{eqnarray}} 
\newcommand{\eea}{\end{eqnarray}}
\newcommand{\be}{\begin{equation}} 
\newcommand{\ee}{\end{equation}}

\newcommand{\lhc}{{\rm lhc}}
\newcommand{\rhc}{{\rm rhc}}

\newcommand{\muD}{m_r}

\newcommand{\mbar}{\bar{m}_D}

\newcommand{\ijs}{\affiliation{Jozef Stefan Institute, Jamova 39, 1000 Ljubljana, Slovenia}}

\newcommand{\msu}{\affiliation{ Department of Physics and Astronomy, Michigan State University, East Lansing, 48824, MI, USA}}

\newcommand{\ulj}{\affiliation{Faculty of Mathematics and Physics, University of Ljubljana, 1000 Ljubljana, Slovenia}}

\def\IMSc{The Institute of Mathematical Sciences, CIT Campus, Chennai, 600113, India}
\def\HBNI{Homi Bhabha National Institute, Training School Complex, Anushaktinagar, Mumbai 400094, India}
\newcommand{\imsc}{\affiliation{\IMSc}}
\newcommand{\hbni}{\affiliation{\HBNI}}

\newcommand{\boc}{\affiliation{Institut für Theoretische Physik II, Ruhr-Universität Bochum, D-44780 Bochum, Germany}}

\newcommand{\sasa}[1]{\textcolor{blue}{#1}}

\begin{document}

 \title{Doubly charm tetraquark  channel  with isospin $1$ from lattice QCD}

\preprint{MSUHEP-24-013}

%%%%%%%%%%%%%%%%%%%%%%%%%%%%%%%%%%%%%%
\author{Lu Meng\orcidlink{0000-0001-9791-7138}}
\email{lu.meng@rub.de}
\boc

\author{Emmanuel Ortiz-Pacheco\orcidlink{0000-0003-4149-1208}}
\email{ortizpac@msu.edu}
\ijs\msu

\author{Vadim Baru\orcidlink{0000-0001-6472-1008}}
% \email{}
\boc

\author{Evgeny Epelbaum\orcidlink{0000-0002-7613-0210}}
% \email{}
\boc

\author{M. Padmanath\orcidlink{0000-0001-6877-7578}}
\email{padmanath@imsc.res.in}
\imsc
\hbni

\author{Sasa Prelovsek\orcidlink{0000-0002-7496-6188}}
\email{sasa.prelovsek@ijs.si}
\ijs \ulj

\begin{abstract} 
Experimentally, the doubly charm tetraquark channel $cc\bar q\bar q$ with $q\!=\!u,d$ features  an exotic hadron, $T_{cc}$,   with isospin $I\!=\!0$ near the $DD^*$ threshold, while no peak  was  observed for  $I\!=\!1$.  We present a lattice QCD study of this channel  with  $I\!=\!1$, $J^P\!=\!1^+$  and $m_\pi\simeq 280~$MeV.  
Finite-volume energies calculated across five charm quark masses consistently   feature a positive energy shift with respect to non-interacting energies, indicating repulsive interaction at energies near threshold. These energies are used to compute the $DD^*$ scattering amplitude using both the standard Lüscher method and the recently proposed effective-field-theory-based approach in the plane-wave basis, which incorporates the  long-range interactions and the left-hand cut. Both analyses render a small negative scattering length and the  scattering amplitude that does not feature any poles in the energy region near the $DD^*$ threshold, in line with LHCb results. We identify that the Wick contraction resembling $t$-channel isovector-vector meson exchanges between $D$ and $D^*$ plays a key role in distinguishing between the $I=0$ and $I=1$ channels,  leading to  repulsion in the $I=1$ and attraction in the $I=0$ channel.

\end{abstract}

\maketitle

\section{Introduction}
\label{sec:introduction}

The LHCb collaboration discovered the doubly charm tetraquark $T_{cc}$ with flavor $cc\bar u \bar d$ and likely spin-parity $J^P=1^+$~\cite{LHCB:2022nat,LHCB:2022natcom}. This state manifests itself
as a very narrow peak in the $D^0D^0\pi^+$ invariant distribution slightly below the $D^0D^{*+}$ threshold, while no narrow signal-like structure was found up to $4~$GeV  in the $D^+D^0\pi^+$ channel related to the $I=1$ channel~\cite{LHCB:2022natcom}. This implies that the observed exotic hadron $T_{cc}$ has isospin $I=0$, while members of the  $I=1$ triplet  $cc\bar u\bar d$, $cc\bar u\bar u$ and $cc\bar d\bar d$  have not been found experimentally. 

This work aims at contrasting the doubly charm tetraquark channels with $I=0$ and $I=1$ theoretically using the framework of lattice QCD. Attraction in the $I=0$ channel was already established in lattice QCD from the $DD^*$ scattering amplitudes~\cite{Padmanath:2022PRL,Chen:2022vpo,Collins:2024PRD,Whyte:2024hep}, the HALQCD potential~\cite{Lyu:2023xro,Ikeda:2013vwa} and the finite-volume energies~\cite{Junnarkar:2018twb,Cheung:2017tnt}. In this channel, $T_{cc}$ appears either as a virtual state pole~\cite{Lyu:2023xro,Padmanath:2022PRL,Chen:2022vpo,Whyte:2024hep} or a resonance below threshold~\cite{Collins:2024PRD,Meng-Lin:2023PRL,Meng:2023bmz} at pion masses heavier than the physical value. 

Lattice studies of the $I\!=\!1$ $DD^*$ channel are more scarce: Positive energy shifts, suggesting a repulsive nature of interaction, were found for finite-volume energies evaluated by CLQCD at $m_\pi\simeq 350~$MeV~\cite{Chen:2022vpo}. This is  qualitatively consistent with the repulsive $DD^*$ potential observed using the HALQCD method at $m_\pi\simeq  410-700~$MeV~\cite{Ikeda:2013vwa}. 
The recent study of  potentials with static heavy quarks found mild attraction at short distances and repulsion at larger distances \cite{Bicudo:2024vxq}. The corresponding solutions of the Schrödinger equation for the physical $D$-meson masses do not support  the possibility of bound states with $I=1$ like all previous static studies  ~\cite{Bicudo:2012qt,Bicudo:2015vta,Bicudo:2015kna}. 
Note that this may not be  applicable directly for $cc\bar u\bar d$, since   corrections to the static approximation are expected to be sizeable in the charm sector. 
Also, it was argued in Ref.~\cite{Baru:2018qkb} that for doubly-heavy molecules,
 sectors with different heavy-quark masses cannot be related within the same heavy-flavor EFT.

\begin{figure*}
        \includegraphics[width=0.90\textwidth]{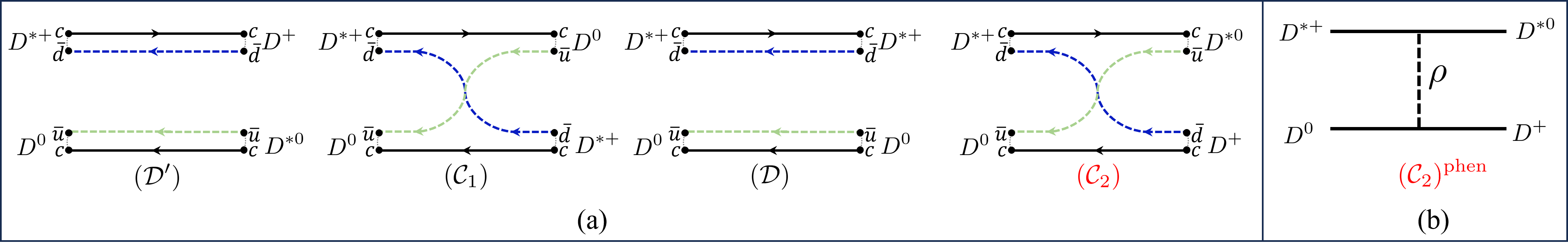}
        \caption{(a) The four Wick quark contractions for $DD^*$ scattering with $I=1$ and $0$. Each contraction is accompanied by a contraction  with the interchange $D^{+}\leftrightarrow D^{0}$ and $D^{*+}\leftrightarrow D^{*0}$; those have equal value and are not shown. (b) The Wick contraction ${\cal C}_2$, which differentiates isospin channels $I=0$ and $I=1$ according to Eq. (\ref{isospins}),  can get contributions, in particular,  from the exchange of a charged $\rho$ meson, while not from $\pi$, $\eta$ or $\omega$ exchanges.    
}
\label{fig:wick}
\end{figure*}

 Let us mention a few possible  reasons  why $I\!=\!0$ channel could be naturally more attractive than $I\!=\!1$:\\
 
 \begin{itemize}
\item  In the diquark-antidiquark scenario, the  light diquark $[\bar u\bar d]_{3_c}$ with the scalar quantum numbers $I(J^P)=0(0^+)$ (also referred to as "good"~\cite{Jaffe:2004ph}) has about $200~$MeV smaller energy than the vector diquark with $I(J^P)=1(1^+)$
(referred to as ``bad")~\cite{Francis:2021vrr,Jaffe:2004ph}. This favors the existence of the exotic state $[QQ]_{\bar 3_c}[\bar u\bar d]_{3_c}$ with $I\!=\!0$ over $I\!=\!1$, at least  
if the heavy quark $Q$ is heavy enough.

\item 
 In the molecular picture, the differences between $I\!=\!0$ and $I\!=\!1$ $DD^*$  channels may seem less obvious
since correlation functions in both isospin channels get contributions from the same set of four quark Wick contractions shown in Fig.~\ref{fig:wick}a and provided in  Sec.~\ref{sec:wick}.
 These contributions combine in specific linear combinations for each isospin, namely
\begin{align}
\label{isospins}
\langle (DD^*)_I| (DD^*)_I\rangle & = {\cal D}-{\cal C}_1+(-1)^{I+1}({\cal D}^\prime-{\cal C}_2)\\
(DD^*)_{I}&=  \tfrac{1}{\sqrt{2}} [ D^0 D^{*+}+ (-1)^{I+1} D^+ D^{*0}]. \nonumber
\end{align}
 As follows from this expression, the terms $\cal D$ and ${\cal C}_1$ stay the same for both isospins. The differences between the two isospin channels result from the opposite signs in front of the Wick contractions ${\cal C}_2$ and $\cal D'$.
Our lattice results, presented in Sec.~\ref{sec:wick}, indicate that the relative contribution from the  ${\cal D}^\prime$ contraction with respect to other contractions is negligible. Thus the   contraction ${\cal C}_2$  plays a key role in determining the nature of interactions in the respective isospin sectors.  These findings agree with   CLQCD~\cite{Chen:2022vpo} and  an extended discussion  is presented in Sec.~\ref{sec:wick}.

To preserve charge conservation, the Wick contraction ${\cal C}_2$  can phenomenologically only get contributions from  isovector meson exchanges.
Moreover, the exchange of a pseudoscalar $P$ meson would require a $DDP$ vertex, which is forbidden by parity conservation. However, the potential for charged $\rho$-meson exchange in this contraction is allowed and is proportional to the isospin factor $\bm{\tau}^{(1)} \cdot \bm{\tau}^{(2)}$, where $\bm{\tau}^{(i)}$ is the Pauli matrix in the vertex $i$
\begin{equation}
\label{rho}
\hspace*{0.5cm}
\langle (DD^*)_I| 
{\bm{\tau}^{(1)} \cdot \bm{\tau}^{(2)}} |(D^*D)_I\rangle_{{\cal C}_2} = \begin{cases} -2\ ;\ {I\!=\!0} \\ 2\ ; \ {I\!=\!1.}\end{cases} 
\end{equation}
To derive this, we used that 
the contribution of $\tau^{(1)}_3\tau^{(2)}_3$   vanishes for  contraction ${\cal C}_2$, see Appendix~\ref{sec:isospin-factors} for details.
As the  isospin factors in Eq.~\eqref{rho} have opposite signs,  
 the exchange of the charged $\rho$-meson may play an important role in distinguishing the two isospin channels within the molecular scenario. 
\end{itemize}

This paper presents a lattice simulation of the $cc\bar u \bar d$  channel with $I=1$ at $m_\pi\simeq 280~$MeV and is an extension of  our earlier simulations of the $I=0$ channel~\cite{Padmanath:2022PRL,Collins:2024PRD}   on the same CLS ensembles. We  extract $DD^*$ scattering amplitude at five values of the charm quark masses and find only very mild dependence on the charm quark mass for $I=1$. The   scattering amplitude is extracted by  employing more than  one lattice volume and considering also non-zero total momenta of the $DD^*$ system.  Our analogous study for $I\!=\!0$ indicated that $T_{cc}$ transitions from a resonance to a virtual state as the charm quark mass is increased~\cite{Collins:2024PRD}. The largest simulated heavy quark mass maybe of relevance\footnote{The reduced mass of the $DD^*$ system  for our Set 5 in Table~\ref{tab:5masses}  is close to the reduced mass of the $B^*D$ system in Nature.} for the channel $bc\bar u\bar d$  that is currently being investigated by LHCb.

 We make significant progress  beyond the only existing  lattice study of the isovector $DD^*$ channel, which was  based on the finite-volume spectra of Ref.~\cite{Chen:2022vpo}.  
 That study   employed the L\"uscher method combined with the effective range approximation,  $p\cot\delta =\tfrac{1}{a_0}+\tfrac{1}{2} r_0 p^2$, and   reported   effective 
 range parameters   that suggested  a bound state pole just below the $DD^*$ threshold. 
 This would imply the existence of an additional finite-volume eigenstate slightly below the threshold,  which, however,  was not observed in the simulation. Moreover,   
 it is important to note  that two major concerns have recently been raised in the context of lattice data analyses.   
 First, the effective-range expansion (ERE) has  a limited range of applicability due to the left-hand cut (lhc) located slightly below the $DD^*$ threshold, 
 which  arises from an on-shell one-pion exchange  in the $u$-channel~\cite{Meng-Lin:2023PRL}.  
 The second concern involves the L\"uscher method itself,  as the recent studies have highlighted challenges for this method when dealing with energy levels below
  left-hand cuts~\cite{Raposo:2023oru,Dawid:2023jrj,Green:2021qol,Meng:2023bmz,Dawid:2024dgy}.  Several extensions and alternative approaches have been proposed 
  to address these issues~\cite{Raposo:2023oru,Meng:2023bmz,Bubna:2024izx,Hansen:2024ffk,Dawid:2024dgy}.  
 While these approaches agree on the necessity of explicitly incorporating the one-pion exchange (OPE), they differ in their technical implementations.  Only a few studies~\cite{Meng:2023bmz,Dawid:2024dgy}  so far applied the framework to actual lattice data. The chiral effective-field theory (EFT) with explicit pions was employed in Ref.~\cite{Meng:2023bmz} to address both issues for isoscalar $DD^*$ scattering in  finite and infinite volumes.

  Following Ref.~\cite{Meng:2023bmz},  we adopt the same EFT approach 
  in this study to analyze $DD^*$ scattering in the isovector channel.   
  Specifically,  the chiral EFT approach relates scattering amplitudes with finite-volume 
  energies across the entire energy region of interest, including along the left-hand cut. We adjust the contact interactions, which parameterize the 
  unknown short-range physics, to best fit our finite-volume energy data. With these parameters fixed from the lattice spectra, we compute and analyze 
  the infinite-volume scattering amplitudes near the $DD^*$ threshold. 
  The plane-wave formulation of the approach allows for a systematic extraction of partial waves contributing to the same finite-volume irreducible representations (irrep)~\cite{Meng:2021uhz}. In this work, it helps in disentangling the contributions from $S$- and $P$-wave scattering in $DD^*$ channel appearing in the $A_2$ irrep in the $P^2=1$ frame.

 In addition,   we utilize the  conventional L\"uscher-based finite-volume quantization conditions in  the partial-wave basis, which are formally valid only above the left-hand cut. Part of the results have already been presented at the proceedings~\cite{Pacheco:2023PoS}. 

The finite-volume lattice energies are presented in the next section and employed to extract the scattering amplitude with two approaches in Sec.~\ref{sec:scat}. The results from scattering analysis are presented and discussed in Sec.~\ref{sec:results}, followed by a discussion of the role and contributions from various Wick contractions in Sec.~\ref{sec:results}. The paper concludes with a brief outlook and conclusions in Section~\ref{Sec:Conc}.

\begin{table*}[htp]
\caption{Scattering length $a_0$ and the effective range $r_0$ for the $DD^*$ system with $I\!=\!1$ for five values of  the heavy quark mass, all at  fixed $m_{\pi} \simeq 280$~MeV.  The results from two approaches are shown, where the chiral EFT
approach includes the one-pion exchange and  employs the cutoff $\Lambda = 1$~GeV, while further results from both approaches are given in Tables~\ref{tab:observables} and~\ref{tab:ERE-observables}. The variation in the heavy quark masses is reflected in  the masses of  $D$ and $D^*$ mesons, $DD^*$ thresholds, reduced mass, the spin average mass $m_{\bar{D}}$ and their corresponding hopping parameter $\kappa_c$ for the bare charm quark mass. Set 2  is closest  to the physical charm quark mass. 
}
\label{tab:5masses}
\begin{ruledtabular}
\begin{tabular}{cccccc}
Set (different $m_c$, all $m_\pi\simeq 280~$MeV) & 1 & 2 & 3 & 4 & 5\\
\hline
$m_D$~[GeV]& 1.762(1)&1.927(1) &2.064(2) &2.191(2) &2.415(2) \\
$m_{D^*}$~[GeV]&1.898(2) &2.049(2) &2.176(2) &2.294(2) &2.506(2) \\
$E_\text{th} =m_D+m_{D^*}$~[GeV]& 3.660(3) & 3.976(3)& 4.240(3) &4.485(3) & 4.922(3) \\
$\mbar =\frac{1}{4}(m_D+3m_{D^*})$~[GeV]&1.864(2) & 2.019(2) & 2.148(2) & 2.269(2) & 2.484(2) \\
$\muD=(m_D^{-1}+m_{D^*}^{-1})^{-1} $~[GeV]& 0.914(1) & 0.993(1) & 1.059(1)& 1.121(1)& 1.230(1) \\
$\kappa_c$& 0.12522& 0.12315& 0.12133145& 0.11956530&0.11627907 \\\hline 
\multicolumn{6}{c}{Effective range expansion}\\
 \hline 
$a_0$~[fm] & -0.39($^{+2}_{-3}$) & -0.37($^{+4}_{-4}$) & -0.37($^{+2}_{-1}$) & -0.38($^{+5}_{-4}$) & -0.40($^{+5}_{-3}$)
 \\\hline 
 \multicolumn{6}{c}{Chiral EFT-based analysis including one-pion exchange}\\
 \hline 
$a_0$~[fm] & -0.31(3) & -0.30(4) & -0.31(3) & -0.31(4) & -0.31(5)\\
 $r_0$~[fm] & -0.87(36) & -0.89(55) & -0.87(36) & -0.87(38) & -0.90(58)
 \end{tabular}
\end{ruledtabular}
\end{table*}

\section{Lattice simulation and eigen-energies }\label{sec:En}

The simulation is performed on two CLS ensembles  with $N_f\!=\!2\!+\!1$ dynamical quarks with $m_\pi\!=\!280(3)$ MeV and lattice spacing $a\! =\! 0.08636(98)(40)$ fm~\cite{Bruno:2014jqa,Bali:2016umi}. The ensembles differ in their spatial volumes $N_L^3=24^3$ and $N_L^3=32^3$,  and they contain 255 and 492 configurations, respectively. The Wilson-clover action is utilized for all quark fields.

The present work investigates    the system  $cc\bar{u}\bar{d}$  with  $I\!=\!1$ for five values of the charm quark masses, where the corresponding $D^{(*)}$ meson masses and spin-averaged mass  $\mbar\!=\!(m_D\!+\!3m_{D^*})/4$ are   provided in Table~\ref{tab:5masses}. 
Set 2 corresponds to the charm quark mass closest to its physical value. 
We  performed an analogous simulation of isospin $I\!=\!0$ channel at same five charm quark masses in Ref.~\cite{Collins:2024PRD}.

The eigen-energies $E_n^{lat}$  are obtained from  the correlation matrices 
\ba
C_{ij}(t)=\langle 0| \mathcal{O}_i(t^\prime + t)\mathcal{O}_j^\dagger(t^\prime)|0 \rangle=\sum_n  Z_i^n Z_j^{*n} e^{-E_n^{lat} t}, \label{correl}
\ea
  via the generalized eigenvalue problem~\cite{Michael:1985ne} using the reference timeslice $t_0=4$. We employ a diverse basis of meson-meson operators $\mathcal{O}$ of the form $D(\vec{p}_1)D^*(\vec{p}_2)$,  where each meson is separately projected to a definite momentum $\vec p_i$
  \begin{eqnarray}
 \label{interpolator-I1-main}
O^{DD^*}_{I=1}
&=&\sum\limits_{k,j}A_{k,j}[D(\vec{p}_{1k})D_j^*(\vec{p}_{2k})]_{I=1}\\
&=& \sum\limits_{k,j}
A_{k,j}[(\bar u\Gamma_{1}c)~(\vec{p}_{1k})~\hspace{0.0cm} (\bar  d \Gamma_{2j}c)~(\vec{p}_{2k})] + \{ u\leftrightarrow d\}.   \nonumber
\end{eqnarray}
Total momenta  $|\vec{P}|=|\vec{p}_1+\vec p_2|=0, 1\frac{2\pi}{L}$ and three irreducible representations in Table~\ref{tab:irreps} are studied in order to  extract $DD^*$ scattering in the partial waves $l=0,1$.  The list of operators  is presented in the Appendix~\ref{app:interpolators}. The quark fields are smeared using the `Distillation' method~\cite{HadronSpectrum:2009krc}, where 60~(90) Laplacian eigenvectors  were employed for $N_L=24$~($32$) ({\it c.f.} Ref.~\cite{Piemonte:2019cbi} for details). The   four Wick contractions that appear in the correlation function \sasa{$C_{ij}$}, Eqs.~(\ref{isospins},\ref{correl}), are shown in Fig.~\ref{fig:wick}a and listed in  Sec.~\ref{sec:wick}.

Diquark-antidiquark $[cc]_{\bar 3_c}[\bar u\bar d]_{3_c}^{I=1}$ interpolator  is not implemented  since the ``bad" light diquark with $I(J^P)=1(1^+)$ has $\simeq 200~$MeV higher energy than the ``good" diquark~\cite{Francis:2021vrr} and is not expected to affect the low-lying spectra. The  operator $D^*(0)D^*(0)$  does not exist for   $I(J^P)=1(1^+)$ due to  Bose symmetry 
while the $D^*D^*$  states with higher discrete momenta lie above the energy region of interest and are therefore not considered. 

\begin{table}[htp]
\caption{The analyzed  total momenta $\vec P$ and irreducible representations, together   with the quantum numbers  characterizing $DD^*$ scattering  ($J,l\leq 2$ are displayed).   }
\label{tab:irreps}
\begin{ruledtabular}
\begin{tabular}{c|c|c|c}
$|\vec P|$ &  Irrep$^P$ & $J^P$ & $\ell$ \\ \hline
$0$ &  $T_1^+$ & $1^+$ & $0,2$\\
$0$ &  $A_1^-$ & $0^-$ & $1$  \\
$\frac{2\pi}{L}\cdot 1$ & $A_2$ & $0^-,1^+,2^-$ & 0,1,2 
\end{tabular}
\end{ruledtabular}
\end{table}

In the non-interacting limit, the energies of the $DD^*$ system  are the sum of the single-meson energies  
\ba
E_{ni}=E_{D(\vec{p}_1)}+E_{D^*(\vec{p}_2)}, \hspace{0.3cm} \vec{p}_i=\vec{n}_i\tfrac{2\pi}{L}, \hspace{0.3cm} \vec{n}_i \in N_L^3,\hspace{0.4cm}  \label{nie}
\ea 
with $E^{con}_{H(\vec{p})}=(m_H^2+\vec{p}^2)^{1/2}$ in the continuum limit. 
  
\begin{figure*}
	\centering
\includegraphics[width=0.30\textwidth]{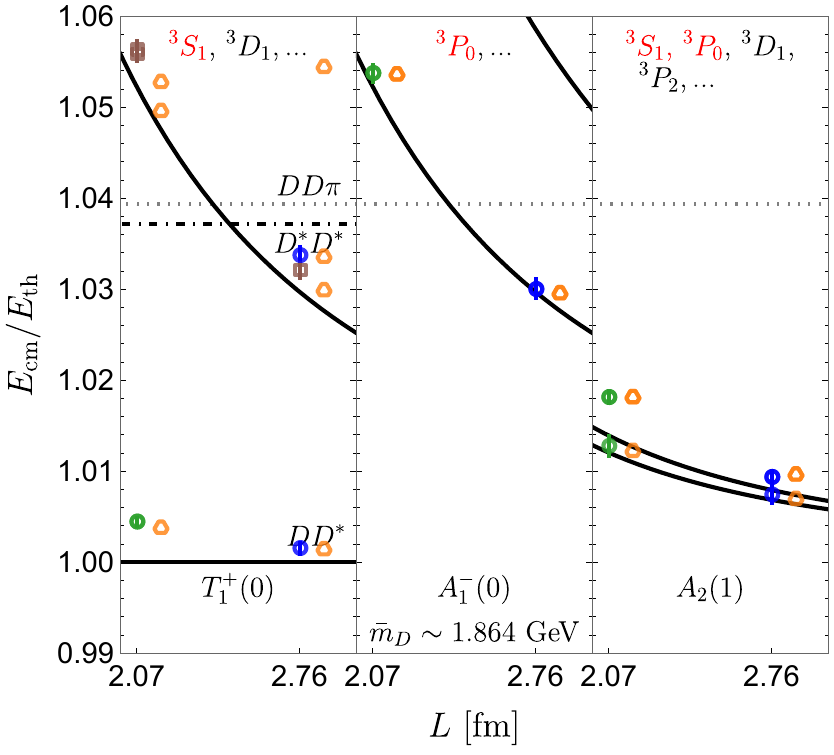}~~~
\includegraphics[width=0.30\textwidth]{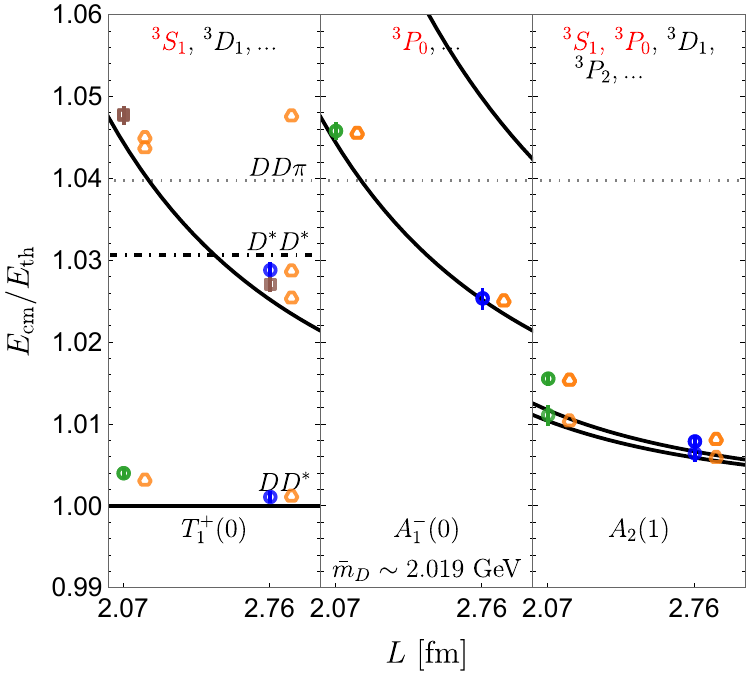}~~~
\includegraphics[width=0.30\textwidth]{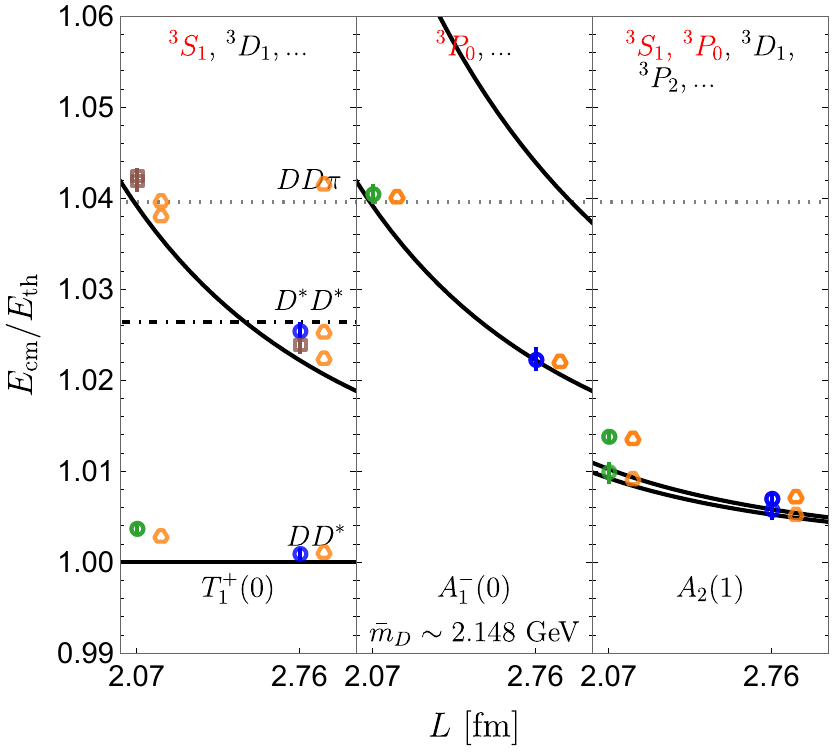}
\includegraphics[width=0.30\textwidth]{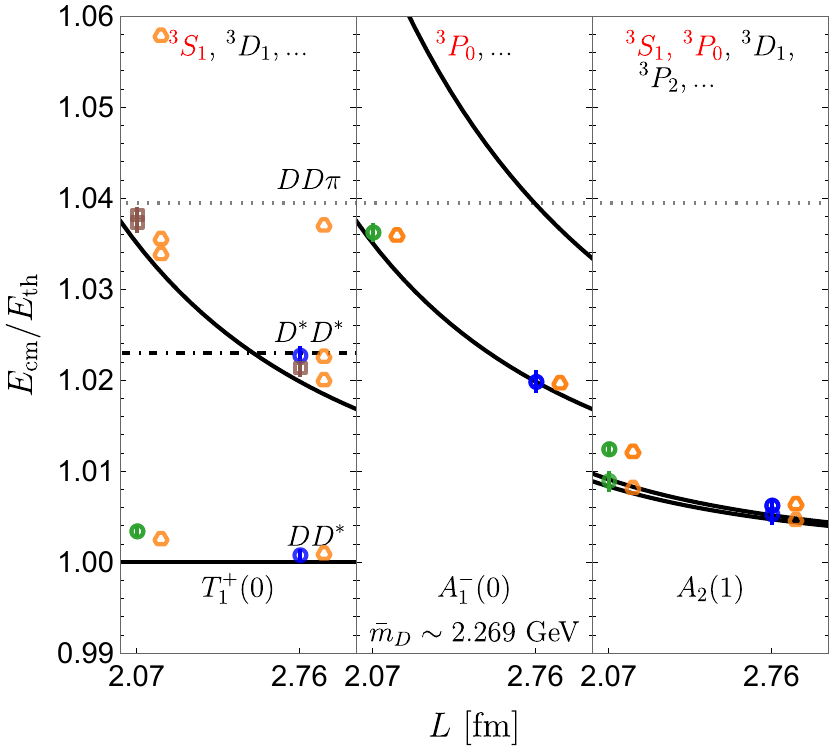}~~~
\includegraphics[width=0.30\textwidth]{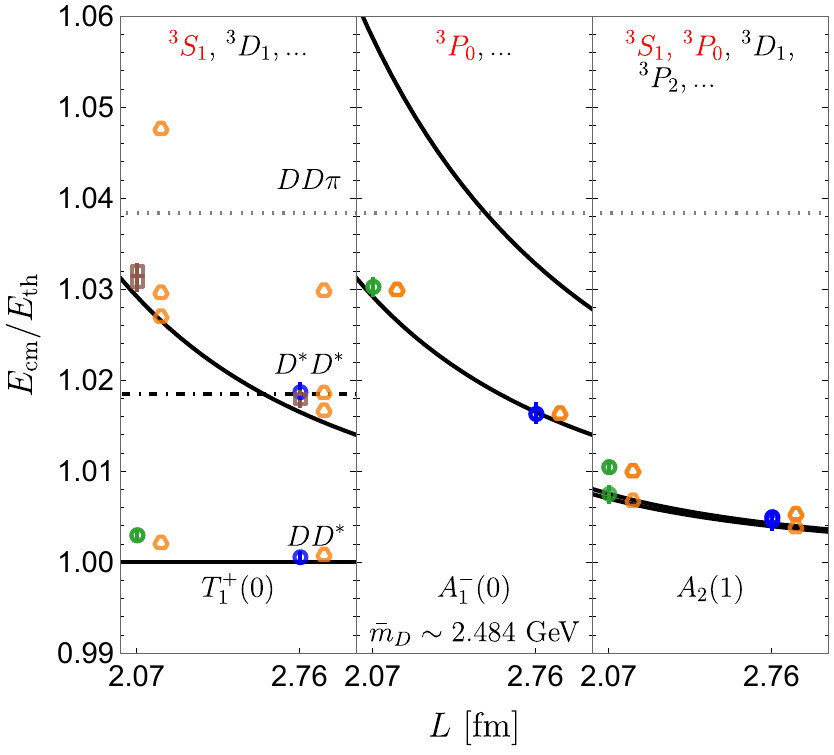}
 	\caption{The finite volume spectra of the  $cc\bar u\bar d$ system with isospin $I=1$. 
  Results for five values of the charm quark mass are shown.
  Lattice energies shown in blue and green are  used in the fits, while  energies shown in brown are omitted from the fits. The lines represent non-interaction energies in Eq.~\eqref{nie}. 
  The orange symbols correspond to the analytically reconstructed energy levels from the chiral EFT approach  including one-pion exchange for the cutoff  $\Lambda=1~$GeV. 
  }\label{fig:spectra_ope_1gev}
\end{figure*} 

The lattice energies of the  interacting system  ${cc\bar{u}\bar{d}}$ with $I=1$ are shown  in Fig.~\ref{fig:spectra_ope_1gev}    for the five  charm quark masses. Each plot contains results for the two volumes, represented by  the blue and green symbols \footnote{$E_{cm}=(E^2-P^2)^{1/2}$ with $E=\Delta E+E_{ni}^{con}$ and $\Delta E =E^{lat}-E_{ni}^{lat}$ are presented to reduce the discretization errors, as  argued in Ref.~\cite{Piemonte:2019cbi}.} (the brown symbols are omitted in the fits), while the continuous curves represent the non-interacting energies (\ref{nie}). Each level dominantly couples to a given type of $DD^*$ operator, which can be inferred from the overlap factors presented in Appendix~\ref{app:interpolators}.

The majority of the energy levels in Fig.~\ref{fig:spectra_ope_1gev}  are shifted upwards with respect to the  corresponding non-interacting $DD^*$ energies, in particular those associated with $DD^*$ in the $S$-wave (from $T_1^+$ and $A_2$). This indicates repulsion between $D$ and $D^*$ in the  channel with $I\!=\!1$ and $\ell\!=\!0$ for all considered charm quark masses. On the other hand, the $DD^*$ interaction in the $P$-wave  seems very small,  as indicated by the negligible shifts of the corresponding levels from $A_1^-$. 
 All five charm quark masses exhibit similar   features of the energy shifts, the only difference being that the $DD^*$ and $D^*D^*$ thresholds approach each other with increasing  the heavy quark mass,   as expected.

\section{Scattering analysis}\label{sec:scat}

The aim is to extract the scattering amplitude  for the single-channel  $DD^*$ system with  $I\!=\!1$ and orbital angular momenta $\ell\!=\!0,1$  in the energy region below the $D^*D^*$ threshold
\ba
T_{\ell}^{(J)}\propto \frac{1}{p \cot\delta_{\ell}^{(J)}-ip}, \hspace{0.2cm} 
\label{amplT}
\ea
where $p\equiv |\vec p|$  is the $D$ and $D^*$ meson momenta in the center-of-momentum frame. 
 In order to extract this from   finite volume energies, we employ two approaches: 
\begin{itemize}
    \item  The main analysis in Sec.~\ref{sec:LQC_PW}  determines the  low-energy constants of the $DD^*$ effective Hamiltonian from  finite-volume energies based on the quantization condition in the plane-wave basis. The EFT potential explicitly includes also the one-pion exchange and thereby incorporates the left-hand cut. The scattering amplitude is determined  based on the resulting potential in the infinite volume.   
\item 
The parameters of the effective range approximation are determined from finite-volume energies also  via the standard L\"uscher approach in Sec.~\ref{sec:LQC_ERE}, that ignores the left-hand non-analyticities.  
\end{itemize}

Both approaches employ the same set of nine finite-volume energy levels, which are shown by green and blue circles in Fig.~\ref{fig:spectra_ope_1gev}. 
We exclude the energy level, that has dominant overlap with the $D(1)D^*(-1)\vert_{l=2}$ interpolator, 
which is the first excited energy level in the $T_1^+$ irrep and the large volume ensemble. 

\subsection{Scattering amplitude based on Lüscher method \label{sec:LQC_ERE}}

Lüscher-based quantization conditions relate the  finite volume energies extracted from  lattice simulations to the partial wave scattering amplitudes.
We parameterize  energy dependence of these partial-wave amplitudes with an effective range expansion (ERE)
\ba
p^{2\ell+1} \cot\delta_\ell^{(J)}=\frac{1}{a_\ell^{(J)}}+\frac12r_\ell^{(J)} p^2.
\label{ERE}
\ea
Here, the parameter $a_\ell^{(J)}$ is the scattering length and $r_\ell^{(J)}$  the effective range which are tuned to ensure that quantization conditions are satisfied for all the considered energy levels at once. 

\begin{figure}[htp]
\includegraphics[height=3.2cm,width=8.3cm]{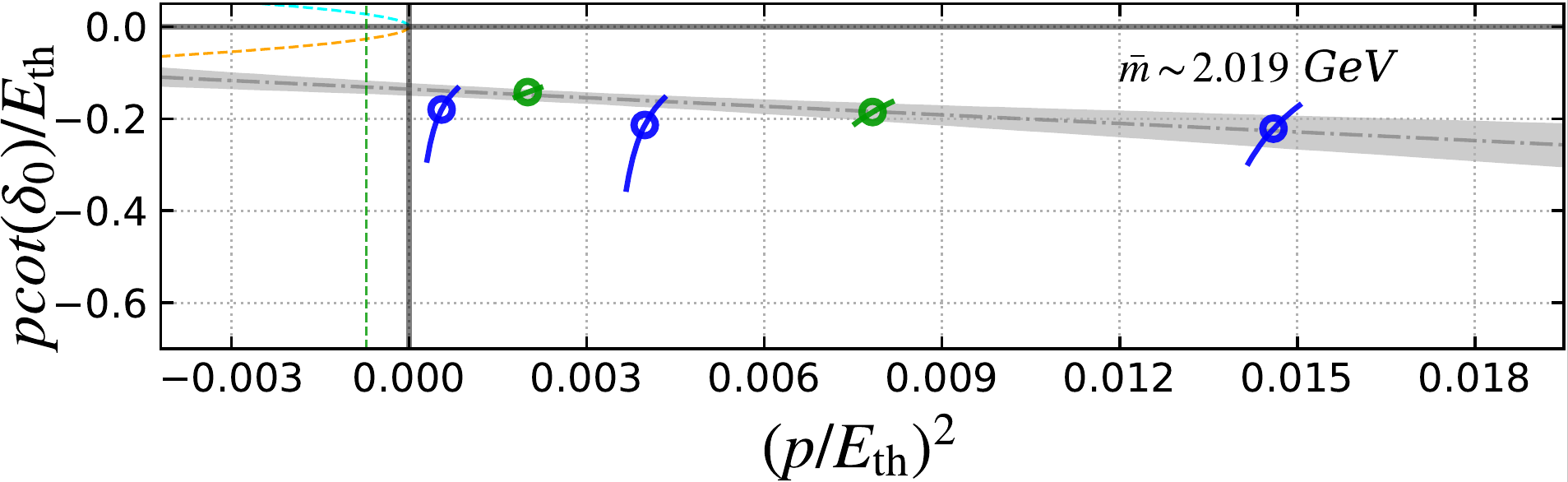}
 \includegraphics[height=3.2cm,width=8.3cm]{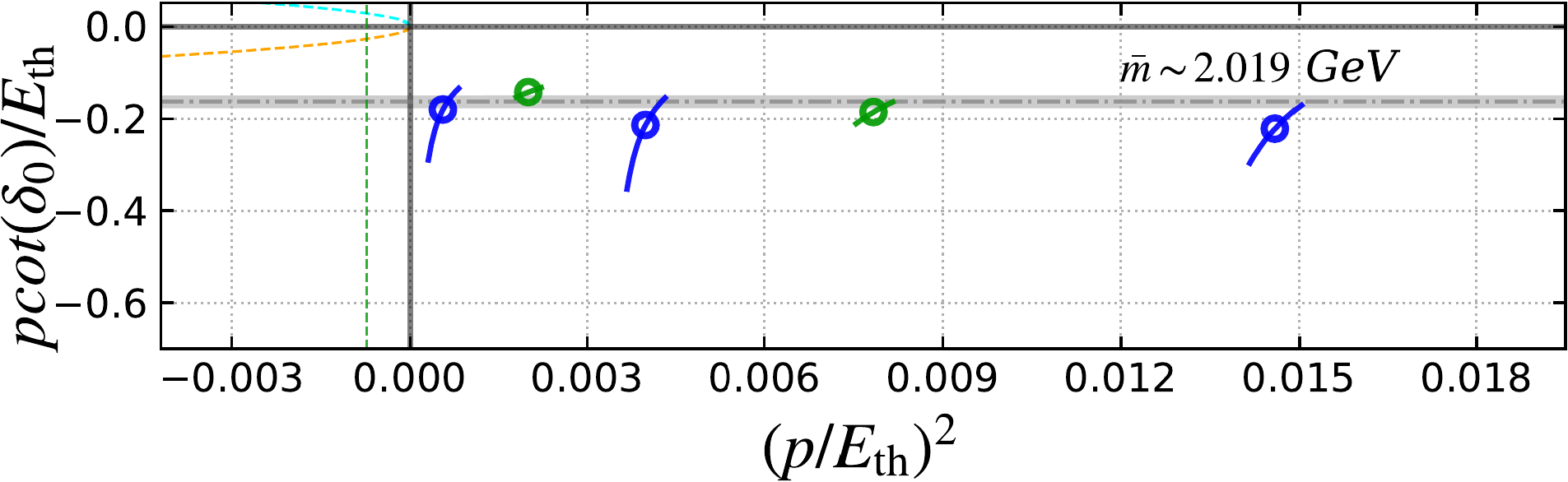}
 	\caption{The momentum dependence of $p\cot\delta$ for    $DD^*$ scattering 
  with $I\!=\!1$ and partial wave $\ell\!=\!0$ at the closest-to-physical charm quark mass (set 2 in Table \ref{tab:5masses}) is represented by the band.    The upper fit,  using an ERE up to ${\cal O}(p^2)$ $p\cot\delta_0=1/a_0+  r_0 p^2/2$, has $\chi^2/\mathrm{dof}=1.6/6$, while the lower fit, using only the leading term $p\cot\delta_0=1/a_0$, also leads to a reasonable $\chi^2/\mathrm{dof}=6.6/7$.   
The bands are obtained from fits, including also the $P$-wave with a constant ERE: $p^3\cot\delta_1=1/a_1$.   The circles represent the simulated finite volume spectra 
  and the corresponding amplitudes are from  L\"uscher's quantization condition, incorporating only the $S$-wave.  Orange and turquoise  dashed lines correspond to 
$ip=\pm |p|$ from unitarity,   normalized to $E_\text{th}$, while the   vertical dashed  line denotes the position of the branch point of the left-hand cut nearest to the threshold. }\label{fig:fitERE}
\end{figure}

The energy dependence of the $\ell=0$ partial wave amplitude in 
the $DD^*$ channel with $I(J^P)=1(1^+)$ is shown in Fig.~\ref{fig:fitERE}.   The two plots in this figure present fits accounting for partial-waves $\ell\!=\!0$ and $\ell\!=\!1$, with a two-parameter ERE for $\ell\!=\!0$ and a one-parameter (constant) ERE for $\ell\!=\!0$. 
Both fits incorporate also the $l=1$ contribution with a one-parameter ERE. 
The best fit parameters from these fits are presented in Appendix~\ref{app:ere}. 
The fits assume negligible interaction 
for $\ell\!=\!2$, as well as insignificant mixing between  
$\ell=0$ and $\ell=2$.  This is supported by the 
fact that  the eigenstate with  dominant overlap to operator $(DD^*)_{\ell=2}$ in $T_1^+$ irrep and $N_L=32$ has a negligible energy shift.  

\subsection{Scattering amplitude based on chiral EFT approach }\label{sec:LQC_PW}

\begin{figure}[htp]
\begin{center}
\includegraphics[width=0.48\textwidth]{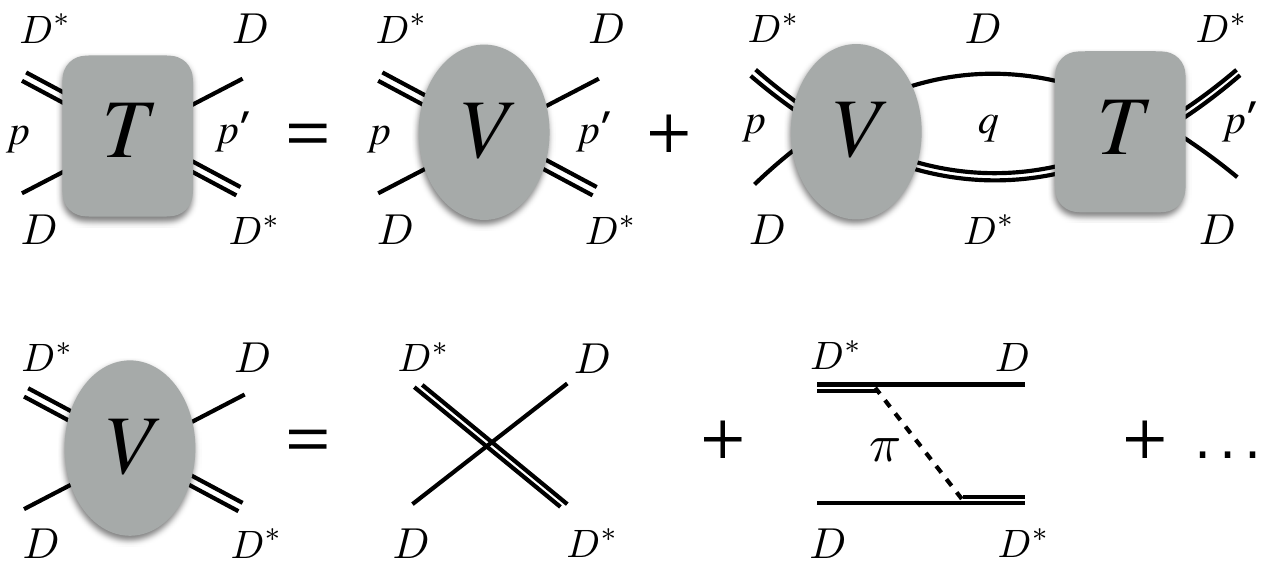}
\caption{\label{fig:LSE}  
Graphical illustration of the Lippmann-Schwinger equations solved in the infinite-volume with the partial-wave projected effective potential truncated to  a given order in the EFT expansion. 
In our calculations, the effective potential consists of the ${\cal O}(Q^2)$ contact interactions from Eqs.~\eqref{eq:Vct} and the one-pion exchange. } 
\end{center}
\end{figure}

To calculate the finite volume (FV) spectra,  we employ the EFT approach that explicitly incorporates long-range  one-pion exchange  and parameterizes short-range physics by a series  of contact interactions.
By adjusting the low-energy constants accompaning contact interactions to achieve the best fit to the lattice FV energy levels, we compute the isovector $DD^*$ scattering amplitude in the infinite volume. 
The infinite-volume scattering amplitude $T$ and the effective potential $V$ are related via the Lippmann-Schwinger equations (LSE), as illustrated in Fig.~\ref{fig:LSE}.  

In finite volume, the LSE takes the matrix form, which can be block-diagonalized according to the lattice irreps \begin{equation}
    \mathbb{T}(E)=\mathbb{V}+\mathbb{V}.\mathbb{G}(E).\mathbb{T}(E)\,,
\end{equation}
with
\begin{eqnarray}
\mathbb{G}(E)=\frac{{\cal J}(\bm{p}_{n})}{L^{3}}G(\bm{p}_{n},E)\delta_{\bm{n}',\bm{n}},
\quad
\mathbb{V}	=V(\bm{p}_{n},\bm{p}_{n'})\,.
\end{eqnarray}
where $\mathbb{V}$ is the effective potential in FV, $\mathcal{J}$ is the Jacobi determinant arising from the transformation between the box  and the center-of-mass  frames,  see Ref.~\cite{Meng:2023bmz} for details,  
 while $p_{\bm{n}}$ and $p_{\bm{n'}}$ are the discretized momenta. 
Further, the $DD^*$ Green function $G$  reads 
\bea
G(p_n,E)=\frac{1}{4\omega_{1}\omega_{2}} \left(  \frac1{E-\omega_1-\omega_2} - \frac1{E +\omega_1+\omega_2} \right),
\eea
where $\omega_i = \sqrt{m_i^2+p_n^2}$  with $m_1 =m_D$  and $m_2=m_{D^*}$.

The effective potential $V$ for $DD^*$ scattering  up to ${\cal O}(Q^2)$ in chiral EFT reads
\begin{equation}
V=V_{\text{OPE}}^{(0)}+V_{\text{cont}}^{(0)}+V_{\text{cont}}^{(2)}+...\ .\label{eq:veft}
\end{equation}
The hierarchy of the operators is based on the expansion parameter $Q \in \{p/\Lambda_b, m_{\pi}/\Lambda_b\}$ with $p$ being a typical soft momentum scale, $m_{\pi}$ the pion mass and $\Lambda_b$ the breakdown scale of the chiral expansion.
In addition, it is assumed that the nonanalytic contribution of the irreducible two-pion exchange 
is suppressed and 
can be well represented by the considered contact interactions.

The most relevant $DD^*$ contact potentials that contribute to the irreps $T_1^+(0), A_1^-(0)$ and $A_2(1)$   to the order ${\cal O}(Q^2)$  are~\cite{Meng:2023bmz}
\bea\nonumber
V_{\text{cont}}^{(0)+(2)}[^3S_1]&=&\left(C^{(0)}_{^{3}S_{1}}  +C^{(2)}_{^{3}S_{1}}  (p^{2}+p'^{2})\right )  ({ \bm\epsilon \cdot \bm\epsilon'^*  }    ) e^{\frac{-(p^n+p'^n)}{\Lambda^n}} \\
    V_{\text{cont}}^{(2)}[^3P_0]&=&C^{(2)}_{^{3}P_{0}}  ({\bm p'\cdot \bm\epsilon'^*}  )  ({\bm p \cdot \bm\epsilon  }  )  e^{\frac{-(p^n+p'^n)}{\Lambda^n}}~\label{eq:Vct},
\eea
where $\bm p$ and $\bm p'$ and $\bm\epsilon$ and $\bm\epsilon'$ denote the cms momenta  and   polarizations. 
The contact interactions 
are supplemented  with the exponential regulators with $n=6$, where  the cutoff $\Lambda$  in what follows is chosen to be either 0.5 GeV or 1 GeV.

The  regularized OPE interaction in the isovector $DD^*$ elastic channel can be written as 
\be\label{Eq:OPE}
V_{\text{OPE}}^{(0)} =  -\frac{m_Dm_{D^{*}}g^2}{f_{\pi}^2} \; \frac{ ({\bm k\cdot \bm\epsilon}  )  ({\bm k \cdot \bm\epsilon'^*  }) }{\bm k^2 + \mu^2} e^{\frac{-(\bm k^2+\mu^2)}{\Lambda^2}},
\ee
where $ \mu^2=m_{\pi}^2-\Delta M^2$,  $\Delta M = m_{D^*}-m_D$ and $\bm k=\bm p'+\bm p$. 
The value of the coupling constant $g$ was extracted in Ref.~\cite{Meng:2023bmz} from  fits  to its physical value and the lattice data of Ref.~\cite{Becirevic:2012pf}. For the given lattice spacing of $a\approx 0.086$ fm and $m_{\pi}=280$ MeV, we found $g=0.517\pm 0.015$~\cite{Meng:2023bmz}. Furthermore, following Refs.~\cite{Meng:2023bmz,Becirevic:2012pf},  we use $f_{\pi}=105.3$ MeV for $m_{\pi}=280$ MeV. 

 When  $p$ and $p'$ are on shell ($p=p'$) and $m_{\pi}{>}\Delta M$,  
the OPE and, consequently, the on-shell $DD^*$ partial wave amplitudes exhibit the left-hand cut
at imaginary values of the momentum.  The lhc branch point closest to the threshold is given by~\cite{Du:2023hlu}
\be
(p_\lhc^{1\pi})^2=-\frac{\mu^2}4=-(126 \ \mbox{MeV})^2\Rightarrow \left(\frac{p_\lhc^{1\pi}}{E_\text{th}}\right)^2\approx-0.001, ~~\label{eq:lhc_num}
\ee
where $E_\text{th}=m_D+m_{D^*}$.  We note that in general,  the OPE may also have a three-body  cut, corresponding to the on shell $DD\pi$  state (for its effect on the $T_{cc}$ properties at the physical pion mass, see Ref.~\cite{Du:2021zzh}).  However, for $m_\pi = 280$~MeV, it starts at momenta far away from the threshold, $p_{\rhc_3}^2=(552 \ \mbox{MeV})^2$~\cite{Du:2023hlu}, which justifies the static approximation used in Eq.~\eqref{Eq:OPE}.

The finite-volume energy levels can be obtained by solving the equation 
\be \label{eq:rel_det0}
\text{det}[\mathbb{G}^{-1}(E)-\mathbb{V}]=0.
\ee
To solve Eq.~\eqref{eq:rel_det0} in a finite volume, we use the plane wave basis instead of  the conventional partial wave expansion, see also Ref.~\cite{Mai:2017bge} for a related discussion.  This facilitates the systematic inclusion of partial wave mixing effects resulting from rotational symmetry breaking in a cubic box~\cite{Meng:2021uhz}.   It should be noticed that contrary to the contact potentials, where only the most relevant interactions contributing to the irreps $T_1^+(0), A_1^-(0)$ and $A_2(1)$ are included,  no partial wave expansion and truncation is made for the OPE  in our plane wave expansion method.

Since the LECs do not depend on the finite volume, once they are extracted from the finite-volume spectra, they can be used to calculate observables in the infinite volume. The infinite-volume $T$-matrix can be obtained by solving the  LSE in the partial wave basis, see Fig.~\ref{fig:LSE} for the graphical illustration, as follows:
\bea
\label{eq:lse}
T_{\alpha\beta}(E,p,p') &=& V_{\alpha\beta}(p,p')\\
& &\hspace{-2.cm} +  
 \int   \frac{{d^3}{q}}{(2\pi)^3} 
V_{\alpha\gamma}(k,q)G(E,q)T_{\gamma\beta}(E,q,p'),\nonumber
\eea
where $\alpha$ and $\beta$ refer to the partial waves. To be more specific, 
the Greek indices either run from 1 to 2 accounting for the $^3S_1$ and $^3D_1$ partial waves, respectively, or  $\alpha= \beta=1$ for the $^3P_0$ case.  The phase shifts then can be obtained from Eq.\eqref{amplT}.

\section{Results and discussion based on the scattering analysis}\label{sec:results}

\begin{figure}[htp]
	\centering
	\includegraphics[height=5.cm,width=8.6cm]{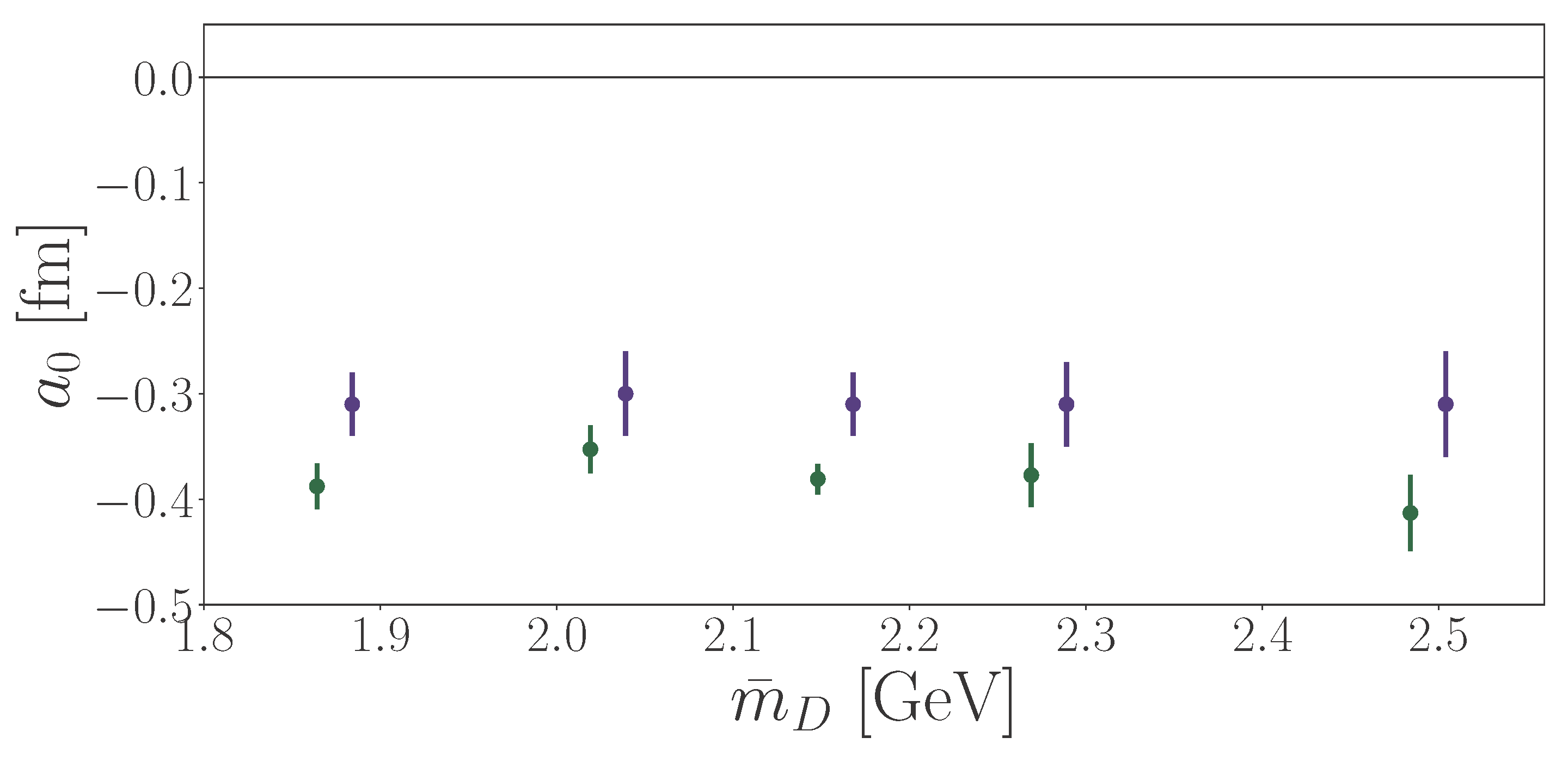} 
 	\caption{ The  scattering length $a_0$ for $DD^*$ scattering with $I=1$ and $l=0$ as a function of the heavy quark mass  in terms of the spin average mass $\mbar =\frac{1}{4}(m_D+3m_{D^*})$.  Comparison between the scattering length values obtained using the EFT approach including one-pion exchange for the cutoff $\Lambda=1$ GeV (violet) and   the Lüscher method, assuming the effective range parameterization and omitting the effect of the left-hand cut (green).  The errors correspond to statistical uncertainties.
  }
  \label{fig:a0}
\end{figure}

A remarkable feature to highlight is that the scattering length $a_0$ in the isospin-1 channel is negative and small for all ERE and EFT fits shown in Fig.~\ref{fig:a0} and  Tables~\ref{tab:5masses},~\ref{tab:observables}  and those in Appendix~\ref{app:ere}.  This contrasts with the isospin-0 channel, where $a_0$ is positive and larger in magnitude, as reported in Refs.~\cite{Padmanath:2022PRL} and~\cite{Meng:2023bmz}.
This appears as a result of the positive energy shifts in the isovector channel, which suggests a repulsive interaction, as opposed to the attractive nature in the isoscalar sector indicated by the negative energy shifts~\cite{Padmanath:2022PRL,Collins:2024PRD}.

\begin{figure*}[htp]
\begin{center}
\includegraphics[width=0.40\textwidth]{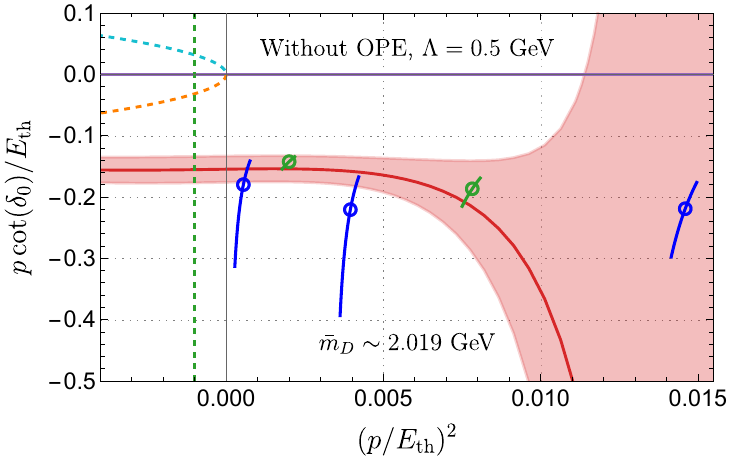}~~~
\includegraphics[width=0.40\textwidth]{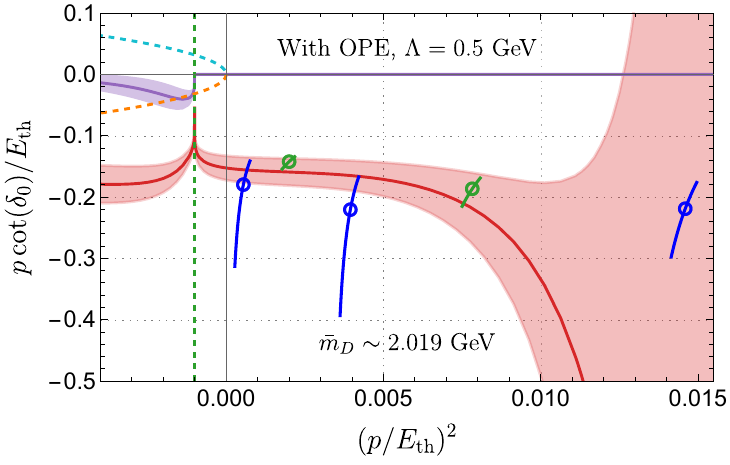}\\
\includegraphics[width=0.41\textwidth]{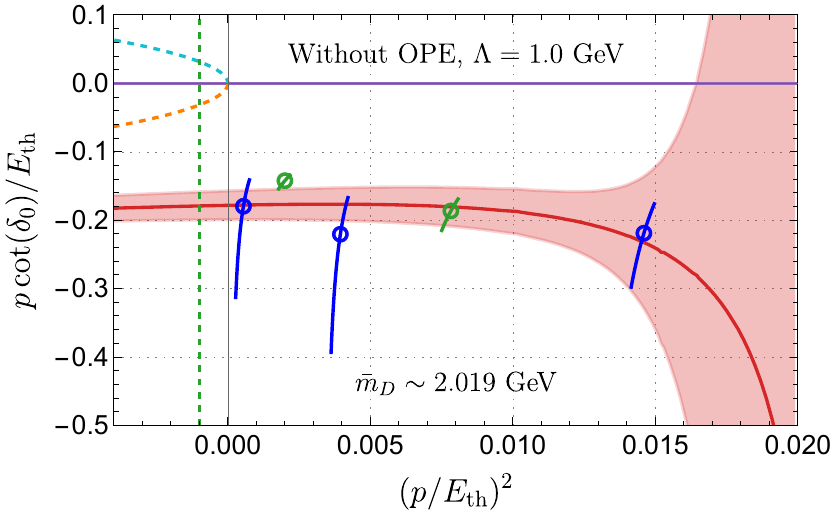}~~~
\includegraphics[width=0.41\textwidth]{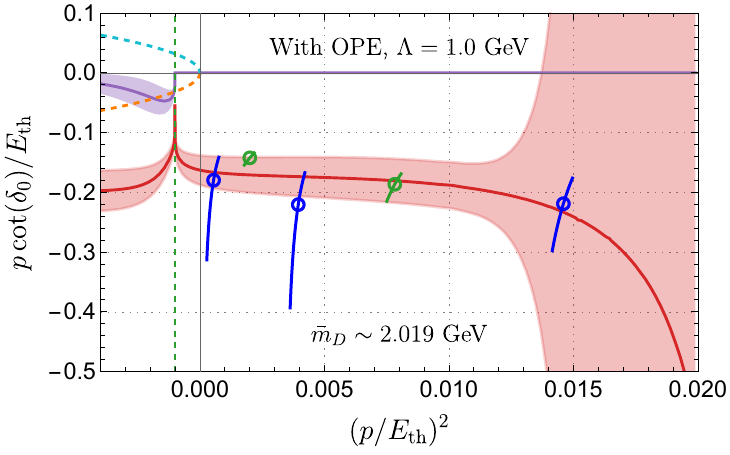}
\caption{\label{fig:phase_shifts_3S1}  
The $p\cot\delta_0$ for $DD^*$ scattering with $I\!=\!1$ and partial wave $\ell\!=\!0$   extracted from the finite-volume energy levels ($E_{FV}$) using the EFT approach without and with the one-pion exchange. The results are shown  for the two cutoffs $\Lambda $   and the $D$-meson masses corresponding to set 2 from Table~\ref{tab:5masses}. 
Red bands represent real part of $ p\cot\delta_0$  from our 3-parameter fits to  $E_{FV}$, including the $1\sigma$ uncertainty. 
The imaginary part of  $ p\cot\delta_0$ shown in violet is nonzero below the  branch point.  
Circles are the   phase shifts extracted from $E_{FV}$ using   L\"uscher's quantization condition   incorporating only  $S-$wave.
 Orange and turquoise  dashed lines correspond to 
$ip=\pm |p|$ from unitarity,   normalized to $E_\text{th}$, while the   vertical dashed  line denotes the position of the branch point of the left-hand cut nearest to the threshold. }
\end{center}
\end{figure*}

In the following, we present the results of the analysis of lattice energy levels ranging from slightly below the $DD^*$ threshold to approximately 100 MeV above it. The highest energy point (see the blue point at the highest energy in Figs.~\ref{fig:fitERE} and~\ref{fig:phase_shifts_3S1}) corresponds to the first excited level in the $T_1^+$ irrep at larger volume, as shown in Fig.~\ref{fig:spectra_ope_1gev}.

 We searched for poles in the resulting infinite-volume scattering amplitude in Eq.~\eqref{amplT} and did not find any in the near-threshold energy region using either of the two methods applied. 
Current experiments also do not observe any hadron structure features in the $DD^*$ final states consistent with an isovector $T_{cc}$ tetraquark, which is in line with our results.

\begin{table*}[tp]
 	\caption{Scattering length $a_0$ and the effective range $r_0$ for the $DD^*$ system with $I\!=\!1$ for five values of  the heavy quark mass, all at  fixed $m_{\pi} \simeq 280$~MeV.  Results are based on chiral EFT with two different cutoffs and including or excluding one-pion exchange. Pole locations (in units of MeV) are provided  relative to the $DD^*$ threshold. 
  All poles lie outside  the region where EFT is applicable and 
   are not visible in Fig.~\ref{fig:eft-all-mc}, since they appear beyond the energy range shown in that figure.  }\label{tab:observables}
   \begin{ruledtabular}
 \begin{tabular}{l|lllll|lllll}
 & \multicolumn{5}{c|}{without OPE, $\Lambda=1$ GeV} & \multicolumn{5}{c}{with OPE, $\Lambda=1$ GeV}\tabularnewline
\hline 
Sets & 1  & 2  & 3  & 4  & 5  & 1  & 2  & 3  & 4  & 5\tabularnewline
\hline 
$\chi^{2}/\text{dof}$  & $4.13/6$  & $3.79/6$  & $3.70/6$  & $3.79/6$  & $4.54/6$  & $2.14/6$  & $2.50/6$  & $2.81/6$  & $3.19/6$  & $4.04/6$\tabularnewline
$a_0~$[fm]  & $-0.28(2)$  & $-0.28(3)$  & $-0.28(3)$  & $-0.28(3)$  & $-0.28(4)$  & $-0.31(3)$  & $-0.30(4)$  & $-0.31(3)$  & $-0.31(4)$  & $-0.31(5)$\tabularnewline
$r_0~$[fm]  & $0.07(6)$  & $0.08(8)$  & $0.09(6)$  & $0.09(8)$  & $0.08(8)$  & $-0.87(36)$  & $-0.89(55)$  & $-0.87(36)$  & $-0.87(38)$  & $-0.90(58)$\tabularnewline
\multirow{2}{*}{pole } & $184(28)$  & $180(34)$  & $169(40)$  & $156(34)$  & $136(32)$  & $187(36)$  & $184(34)$  & $169(40)$  & $158(40)$  & $136(31)$\tabularnewline
 & $-i46(19)$  & $-i47(23)$  & $-i45(22)$  & $-i39(24)$  & $-i33(21)$  & $-i50(29)$  & $-i51(26)$  & $-i45(22)$  & $-i42(22)$  & $-i35(20)$\tabularnewline
\hline 
 & \multicolumn{5}{c|}{without OPE, $\Lambda=0.5$ GeV} & \multicolumn{5}{c}{with OPE, $\Lambda=0.5$ GeV}\tabularnewline
\hline 
$\chi^{2}/\text{dof}$  & $1.86/4$  & $2.08/4$  & $2.18/4$  & $2.53/4$  & $4.03/4$  & $1.58/4$  & $2.10/4$  & $2.40/4$  & $2.87/4$  & $4.29/4$\tabularnewline
$a_0~$[fm]  & $-0.33(4)$  & $-0.32(4)$  & $-0.33(4)$  & $-0.33(6)$  & $-0.32(8)$  & $-0.33(3)$  & $-0.33(4)$  & $-0.33(5)$  & $-0.33(5)$  & $-0.33(7)$\tabularnewline
$r_{0}~$[fm]  & $0.07(16)$  & $0.07(18)$  & $0.09(23)$  & $0.09(18)$  & $0.09(108)$  & $-0.69(32)$  & $-0.73(31)$  & $-0.69(42)$  & $-0.70(41)$  & $-0.77(49)$\tabularnewline
\multirow{2}{*}{ pole } & $90(12)$  & $84(11)$  & $78(11)$  & $75(10)$  & $68(12)$  & $92(11)$  & $86(9)$  & $80(10)$  & $77(7)$  & $70(10)$\tabularnewline
 & $-i47(5)$  & $-i44(5)$  & $-i41(4)$  & $-i39(7)$  & $-i36(5)$  & $-i50(6)$  & $-i47(4)$  & $-i43(6)$  & $-i41(5)$  & $-i38(6)$\tabularnewline
\end{tabular}
\end{ruledtabular}
\end{table*}

\begin{figure}[htp]
    \centering
    \includegraphics[width=1.0\linewidth]{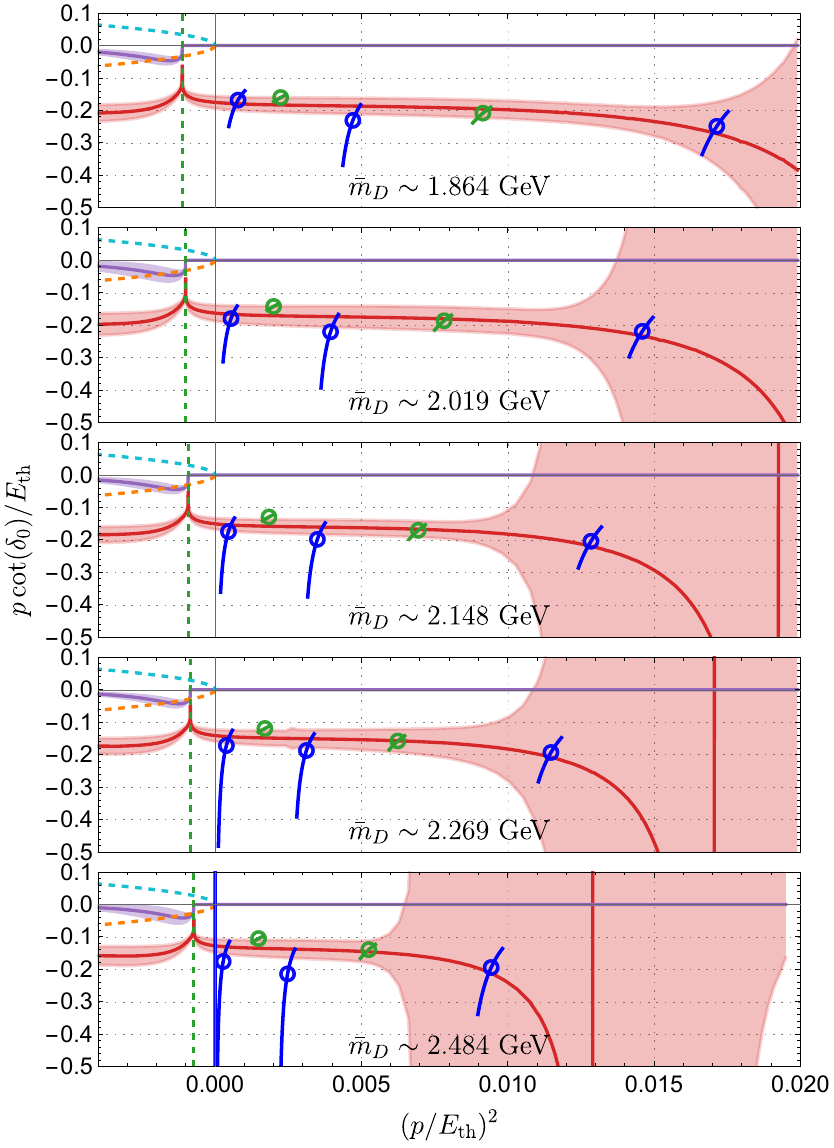}
    \caption{The function $p\cot\delta_0$ for $DD^*$ scattering with $I\!=\!1$ and partial wave $\ell\!=\!0$   for five charm quark masses. The results are extracted   using the EFT approach  including the one-pion exchange for the cutoff $\Lambda=1~$GeV.  See caption of Fig.~\ref{fig:phase_shifts_3S1} for more information. }
    \label{fig:eft-all-mc}
\end{figure}

  \begin{figure*}[t]
\begin{center}
\includegraphics[width=0.7\textwidth]{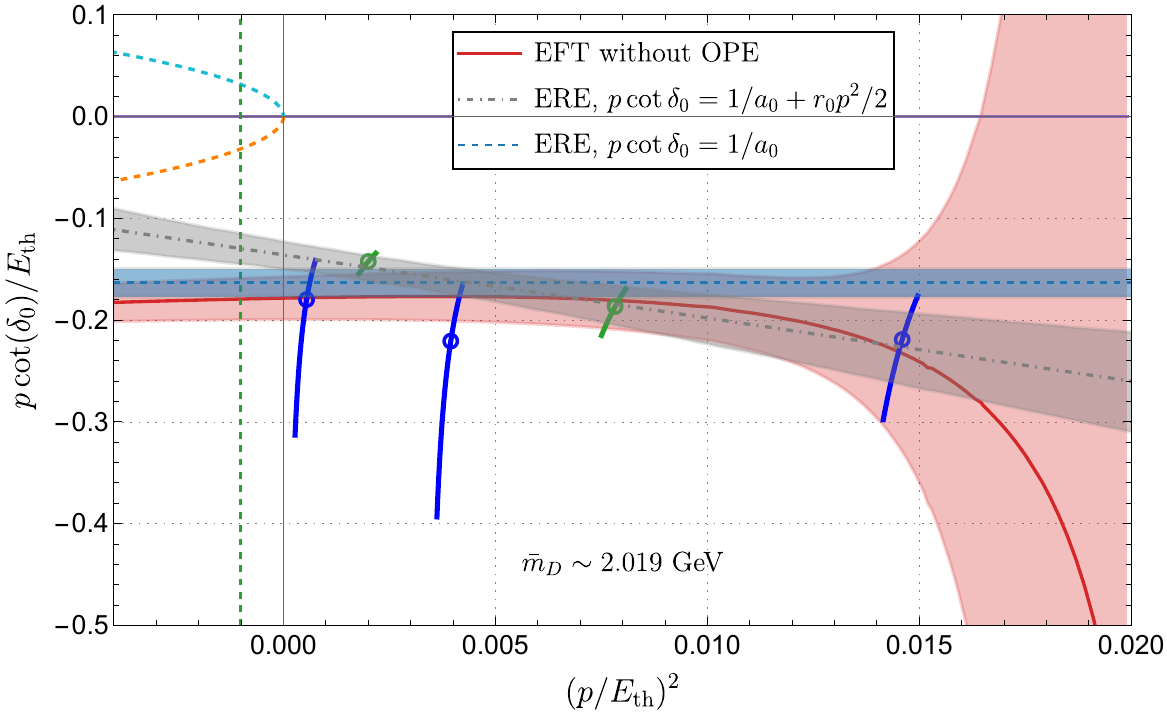}~~~
\caption{\label{fig:phase_shifts_3S1_add}  
The function $p\cot\delta_0$ for $DD^*$ scattering with $I\!=\!1$ and partial wave $\ell\!=\!0$. The plot compares results based on the contact EFT approach without OPE, evaluated with a cutoff $\Lambda=1~$GeV (already shown at bottom left in Fig.~\ref{fig:phase_shifts_3S1}), and two fits based on the effective range expansion  (already shown in Fig.~\ref{fig:fitERE}). More information is given in the captions of these two figures.     }
\end{center}
\end{figure*}

The fit results for the closest-to-physical charm quark mass are shown in Figs.~\ref{fig:phase_shifts_3S1} and~\ref{fig:fitERE}  for the EFT and   ERE approaches, respectively, while Fig.~\ref{fig:eft-all-mc} and Tables~\ref{tab:observables} and Appendix~\ref{app:ere} provide results for all charm quark masses.  The results of the EFT analysis are shown in Fig.~\ref{fig:phase_shifts_3S1}  for the two cutoff values $\Lambda=0.5$ GeV (left panel ) and $1$ GeV (right panel). 
  Please note that  the cutoff $\Lambda=0.5$ GeV is below the highest-energy lattice point (blue dot), which is therefore ignored in the analysis for this cutoff.  All lattice data points are included in fits for the cutoff $\Lambda=1$ GeV.

 Here, several remarks are in order:
 
 \begin{enumerate}[label=(\alph*)] 
 
\item The  left hand cut reveals itself in the discontinuity of  $p \cot (\delta_0)$ at the lhc branch point, below which it develops an imaginary part;  
\item 
As already mentioned earlier, the $DD^*$ scattering length  extracted from the lattice spectra remains   stable against the inclusion of the long-range dynamics from the OPE and variations in the cutoff. 
 The observed small negative scattering length implies a repulsive interaction, given the definition of the ERE in Eq.~\eqref{ERE}. 
\item  The effect of the lhc on $DD^*$ scattering in the isovector channel is less pronounced than in the isoscalar channel, which is consistent with the suppression of the isovector isospin coefficients in the OPE, as compared to their isoscalar counterparts: 
\begin{equation}
\label{Eq:isoOPE}
\langle (DD^*)_I| 
{\bm{\tau}^{(1)} \cdot \bm{\tau}^{(2)}} |(D^*D)_I\rangle_{\rm OPE} = \begin{cases} 3\ ;\ {I\!=\!0} \\ 1\ ; \ {I\!=\!1}\end{cases} 
\end{equation}
Nevertheless, the lhc noticeably impacts the slope of $p \cot (\delta_0)$ near the threshold, influencing the magnitude of the effective range.  Specifically, while the effective range is consistent with zero in a pure contact theory, it becomes slightly negative (around $-1$ fm) when the OPE is included;  
\item  The EFT analysis shows that $p \cot (\delta_0)$ exhibits a pole,
 which corresponds to a zero in the scattering amplitude, at relatively high momenta above the threshold. This pole appears above the maximum energy point used in fits.
This leads to a significant increase in the uncertainty of our calculations at higher momenta. Furthermore, the $S-$wave scattering amplitude also contains 
a resonance pole located even further from the threshold; see Table~\ref{tab:observables} for its position.  
 Importantly, this resonance pole lies beyond the  applicability range of the EFT. 
Therefore, it is reassuring to note that neither the zero nor the pole 
 in the scattering amplitude affect low-energy observables, such as the effective-range parameters, extracted near the $DD^*$ threshold.

\item  
The EFT approach does not lead to any poles below the threshold.
If the ERE approach were applicable below the left-hand cut, the intersection of $p\cot\delta_0$ and $ip$ would imply a bound state and, consequently, an additional energy level that is not observed in the simulation.  This intersection occurs more than $40~$MeV below the threshold for fits based on $p\cot\delta=1/a_0+r_0p^2/2$ and even further below for fits using $p\cot\delta=1/a_0$. In both cases this is far below the branch point of the left-hand cut, where the effective range expansion is no longer valid.  Notably, when comparing our ERE results with those in Ref.~\cite{Chen:2022vpo}, we observe that the effective range in the isovector channel of Ref.~\cite{Chen:2022vpo} is significantly larger. This leads to a steeper slope in the ERE fit, a   bound-state pole  only $-16~$MeV below threshold (almost overlapping with  the  branch point of the left-hand cut)  which would lead to a subthreshold finite-volume level that is not  observed  in Ref.~\cite{Chen:2022vpo}.

To further explore this issue and estimate the impact of potential systematic errors, we provide the results for the ERE fit with $r_0 = 0$ (represented by a straight horizontal line) in Fig.~\ref{fig:fitERE}; see also Fig.~\ref{fig:phase_shifts_3S1_add} for comparison with other fits. This fit still achieves a $\chi^2/\text{dof} \approx 1$ and appears in line with the contact EFT results (as shown in the left panel of Fig.~\ref{fig:phase_shifts_3S1}). Thus, we believe that both approaches can yield consistent results, provided systematic errors are carefully estimated and included. In any case, as discussed above, in the full EFT analysis with pions, the effective range is primarily driven by the rapid variation of  $p \cot \delta$ near the lhc branch point--an effect neglected in both the ERE and contact EFT analyses, which leads to a slightly negative effective range (see Table~\ref{tab:5masses} for the results). 

\begin{figure*}
    \centering
    \includegraphics[width=0.4\textwidth]{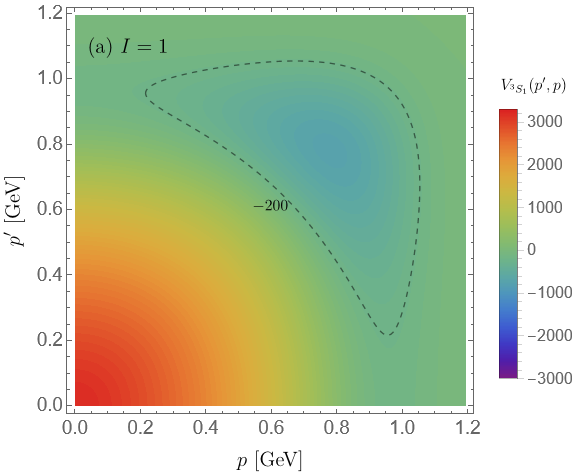}
    ~~~~~~~~~~~~~~~~~~~~~~~~~~
    \includegraphics[width=0.4\textwidth]{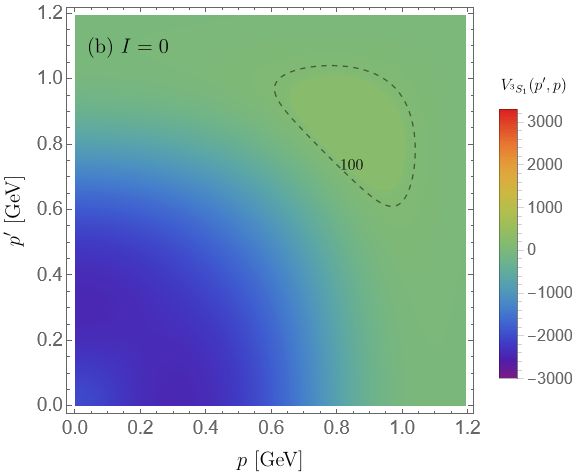}
    \caption{Potentials for the $DD^*$ interactions in the $I=1$ (left panel) and $I=0$ (right panel) channels. To show the negative region in the $I=1$ channel and the positive region in the $I=0$ channel, the dashed contours corresponding to $V=-200$ and $V=100$ are shown in the left panel and right panel, respectively.}
    \label{fig:ptl}
\end{figure*}

\item 
The interactions determined within the chiral EFT approach are shown in Fig.~\ref{fig:ptl}, which also includes the potential for the $I=0$ channel~\cite{Meng:2024PRD} for comparison. The $I=1$ potential is predominantly positive, with a small negative region, whereas the $I=0$ potential is mostly negative, with a minor positive region. However, it is important to note that the potential itself is not an observable. Therefore, one should avoid comparing potentials constructed under different assumptions. In principle, potentials related by unitary transformations can exhibit entirely different features while yielding identical observable predictions~\cite{Bogner:2009bt}.  Consequently, while it is valid to compare the potentials determined within chiral EFT for $I=0$ and $I=1$,  comparing the nonlocal potentials derived in this framework with those based on local assumptions is meaningless. 

\item 
No reliable results for the $P-$wave low-energy parameters can be extracted from the energy levels in $A_1^-$ shown in Fig.~\ref{fig:spectra_ope_1gev}. Indeed, while the lowest energy level (blue point) coincides with the non-interacting energy, offering no useful constraints, the green point corresponds to a high momentum ($\approx 600$ MeV), which alone cannot constrain the low-energy parameters.

  \end{enumerate}

\section{Wick contractions}\label{sec:wick}

\begin{figure*}[htp]
	\centering
\includegraphics[height=6.8cm,width=8.91cm]{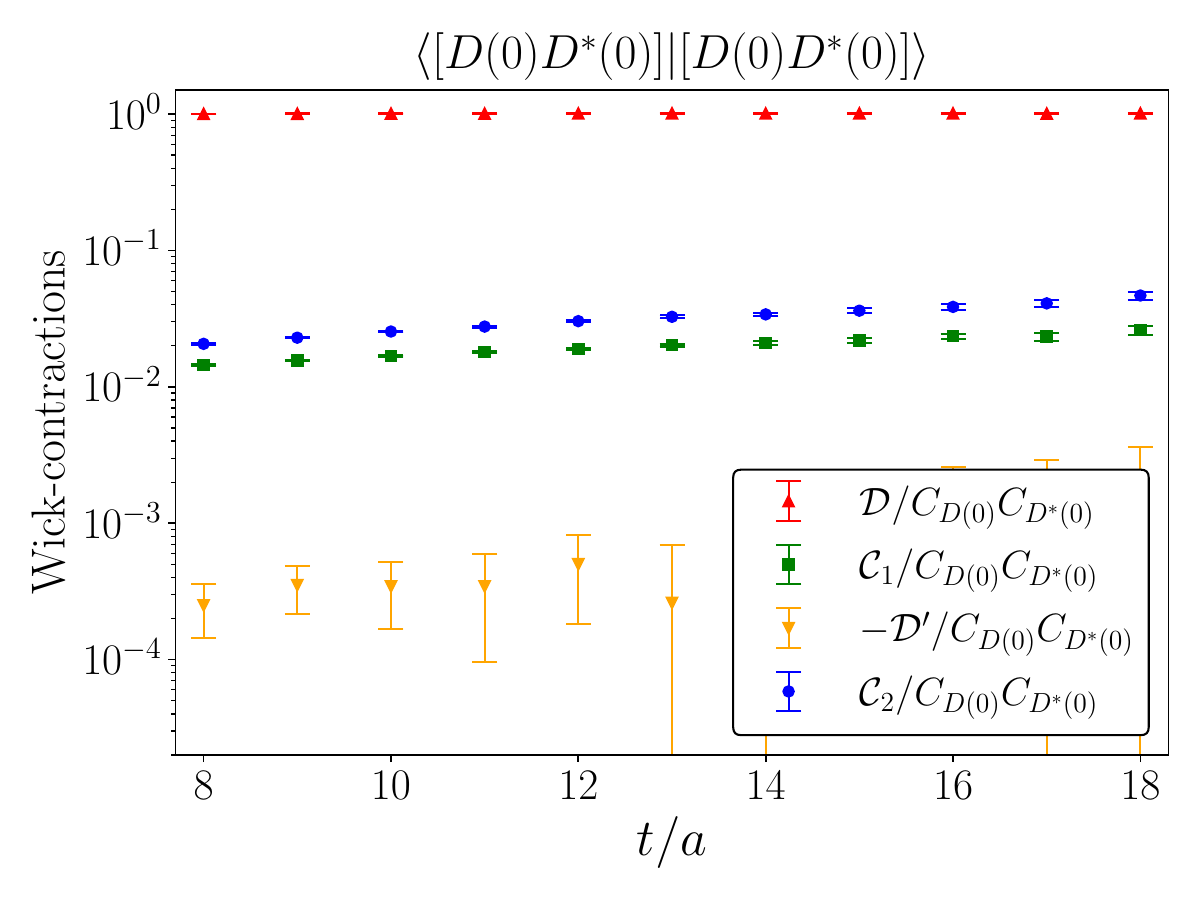}
\includegraphics[height=6.8cm,width=8.91cm]{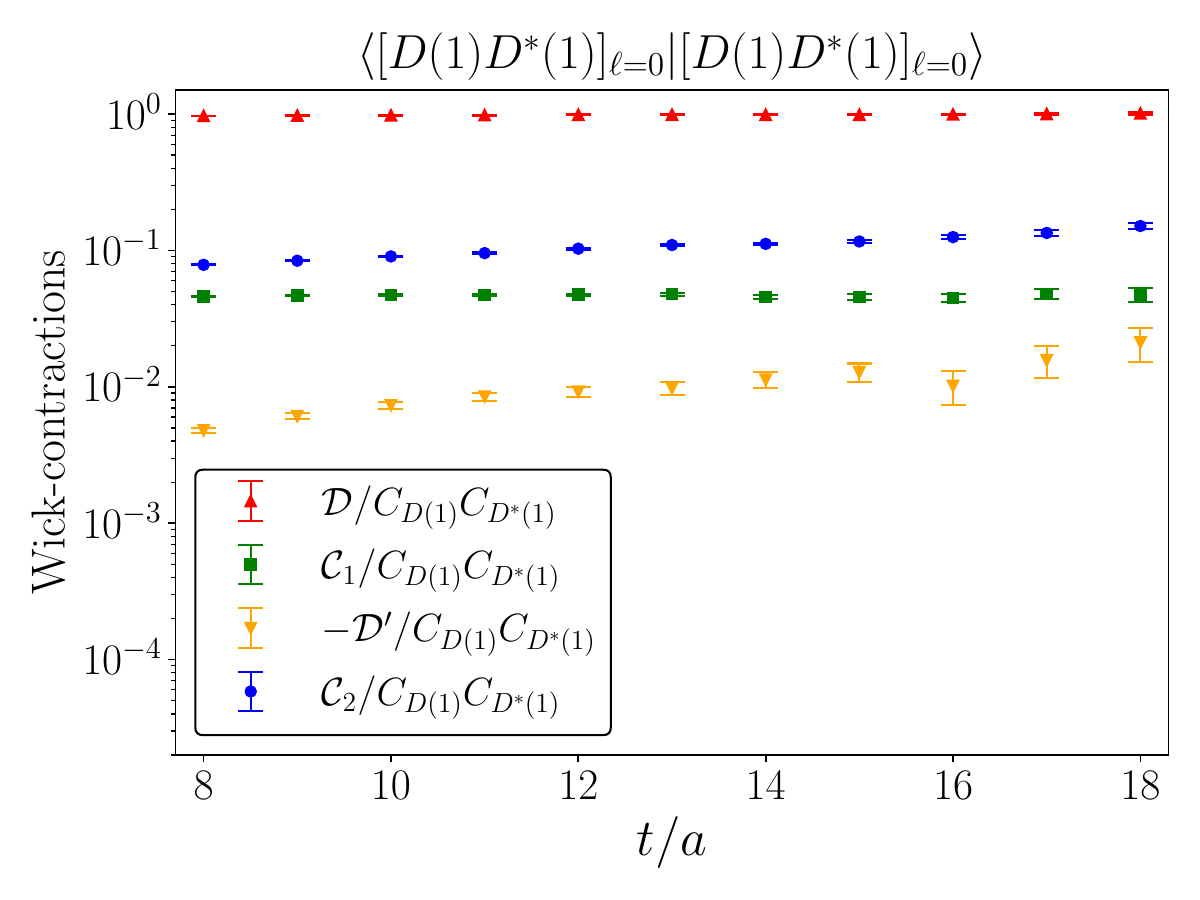}
 \caption{Contribution of  four Wick contractions to the  correlation functions of the $DD^*$ system with $J^P=1^+$. The corresponding diagrams are displayed in Fig.~\ref{fig:wick}. Both plots correspond to diagonal correlation functions $C_{ii}=\langle O_i O_i^\dagger\rangle$ for total momentum $\vec P=\vec 0$ and irreducible representation $T_1^+$, where the left  plot   is for  the operator $O^{D(0)D^*(0)}$ and the right  one is for $O^{D(1)D^*(1)}|_{\ell=0}$   (Table~\ref{tab:operators}). The denominators in the ratios are $\langle C_{D}\rangle \langle C_{D^*}\rangle$, where each of the two correlation functions is separately averaged over gauge configurations. The result is shown for $N_L=32$, while the one for  $N_L=24$ is very similar (the later was provided in \cite{Pacheco:2023PoS} with a normalization that differs by factor 1/2).
 }\label{fig:wick-terms}
\end{figure*}

 The lattice simulations of $cc\bar u\bar d$ channel with $J^P=1^+$ for $I=1$  here    and for $I=0$ in previous studies~\cite{Padmanath:2022PRL,Chen:2022vpo,Collins:2024PRD,Whyte:2024hep} employ only $DD^*$-type interpolators (see Appendix~\ref{app:interpolators}), where each meson is separately projected to a definite momentum. Then all correlation functions $C_{ij}$ are a sum of four Wick contractions, 
 \begin{align}
\label{isospins-sec}
\langle (DD^*)_I| (DD^*)_I\rangle & = {\cal D}-{\cal C}_1+(-1)^{I+1}({\cal D}^\prime-{\cal C}_2)
\end{align}
presented already in Equation  (\ref{isospins}) and in Fig.~\ref{fig:wick}. 
 The Wick contractions are appropriate traces of quark propagators ${\cal S}^{-1}$ and gamma matrices for  pseudoscalar  $\Gamma_P= \gamma_5,  \gamma_5\gamma_t$ and vector $\Gamma_V =\gamma_i,\gamma_i\gamma_t$ mesons 
 \begin{align}
\cal D&=\mathrm{ Tr}[{\cal S}^{-1}_c \Gamma_V {\cal S}^{-1}_q \Gamma_V]\mathrm{ Tr}[{\cal S}^{-1}_c \Gamma_P {\cal S}^{-1}_q \Gamma_P]\\
\cal D^\prime &=\mathrm{ Tr}[{\cal S}^{-1}_c \Gamma_V {\cal S}^{-1}_q \Gamma_P] \mathrm{Tr}[{\cal S}^{-1}_c \Gamma_5 {\cal S}^{-1}_q \Gamma_V]\nonumber\\
{\cal C}_1&= \mathrm{Tr}[{\cal S}^{-1}_c \Gamma_V {\cal S}^{-1}_q \Gamma_V {\cal S}^{-1}_c \Gamma_5 {\cal S}^{-1}_q \Gamma_5]\nonumber\\
{\cal C}_2&= \mathrm{Tr}[{\cal S}^{-1}_c \Gamma_V {\cal S}^{-1}_q \Gamma_P {\cal S}^{-1}_c \Gamma_5 {\cal S}^{-1}_q \Gamma_V]~,\nonumber
\end{align}
where  space-time  positions and momentum projections are omitted for simplicity. 

The Wick contractions  that contribute to  diagonal  correlation functions $C_{ii}=\langle O_i(t) O_i(0)\rangle$   relevant for system with $J^P=1^+$ and total momentum zero are plotted in Fig.~\ref{fig:wick-terms}. The relative contribution of the contraction $\cal D^\prime$ is orders of magnitude too small to have  observable effects on the total correlation function, as can be readily observed from the orange points in Fig.~\ref{fig:wick-terms}. Therefore the main difference between the isospin channels $I=0$ and $I=1$ in Eq. (\ref{isospins-sec}) has to arise from the Wick contraction ${\cal C}_2$, which emerges together with a sign factor that depends on the isospin channel. The  contractions $\cal D$ and ${\cal C}_1$ have the same contribution to  both isospin channels.

\begin{figure}[h]
	\centering
\includegraphics[height=2cm ]{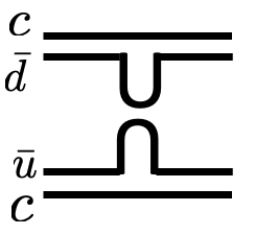}
 \caption{Sketch of a possible   $\rho^0$ or $\omega$ exchange within the Wick contraction ${\cal D} $.
 }\label{fig:wick-D}
\end{figure}

 Let us examine phenomenologically which meson  exchanges may contribute to the four Wick contractions, focusing, in particular,  on single-meson exchanges of light pseudoscalars ($\pi$ and $\eta$) and vectors ($\rho$ and $\omega$):
\begin{itemize}
\item $\cal D$: This contraction is largest as it is nonzero even in the limit when $D^*$ and $D$ do not interact. 
Pseudoscalar exchanges do not contribute here, as a $DDP$ vertex is prohibited by parity conservation.
  The neutral $\rho_0$ and $\omega$ can be exchanged as sketched in Fig.~\ref{fig:wick-D}.  However,  assuming SU(3) symmetry, these two contributions cancel each other. 
  \item $\cal D^\prime$: Neutral $\pi,~\eta,~\rho,~\omega$ mesons can be exchanged.  However, similar to $\cal D$,  the combined $\rho+\omega$  contribution vanishes in the  SU(3) limit. Also, the  $\pi$ and $\eta$ contributions would 
  exactly cancel   each other  in this limit. Due to SU(3) symmetry breaking, however, their masses are not identical, leading to a small but non-vanishing contribution. 
\item ${\cal C}_2$: Charged $\rho^\pm$ mesons can be exchanged, as already discussed in the Introduction. 
The  pseudoscalar exchanges are prohibited by parity conservation  and neutral vector exchange  is prohibited by charge conservation in each vertex.  
\item ${\cal C}_1$: Charged $\pi^\pm$ and $\rho^\pm$ mesons can be exchanged here.  
However, the $\rho^\pm$ contribution in ${\cal C}_1$ differs from that in ${\cal C}_2$, as the vertices involved in these operators come from two different $D^{(*)}D^{(*)}V$ Lagrangian densities: 
 ${\cal C}_1$ contains derivative $D^{(*)}D^{(*)}V$ vertices, 
while  ${\cal C}_2$ is  controlled  by  vertices without derivatives~\cite{Casalbuoni:1996pg}. 
Since these contributions involve different couplings, no model-independent comparison of ${\cal C}_1$ and ${\cal C}_2$ is  possible.
\end{itemize}  

In addition to the mesons exchanges  listed above,  other contributions are possible such as, e.g.,  scalar meson exchanges.  

In conclusion, the Wick contractions ${\cal C}_2$ plays a crucial role in distinguishing between isospins $I=0$ and $1$, with contributions specifically arising from light-meson exchanges with isovector-vector quantum numbers. 
Phenomenologically, this could correspond to $\rho$-meson exchange, potentially significant for the existence of
 the $T_{cc}$ with $I=0$ (see also  Ref.~\cite{Gil-Dominguez:2024zmr} for further phenomenological discussion). Within chiral EFT, this highlights the potential importance of isovector-vector correlated two-pion exchange contributions.
 We note in this context that the important role of two-pion exchange in the $DD^*$ potential at separations $r\simeq 1-2~$fm was highlighted in Ref.~\cite{Lyu:2023xro} at nearly physical $m_\pi$, (while the isospin and total spin of the $\pi\pi$ system were not identified).

\section{Conclusions}\label{Sec:Conc}

We performed a   lattice simulation of the doubly charmed tetraquark channel  $cc\bar{u}\bar{d}$  with isospin $I\!=\!1$ and  $J^P\!=\!1^+$ for five   charm quark masses at fixed $m_\pi\!\simeq\! 280~$MeV. 
The extracted $DD^*$ scattering amplitude near threshold indicates repulsion, with a small negative scattering length $a_0$, which is almost independent of the charm quark mass. No poles are found in the  scattering amplitude  in the relevant energy region near the $DD^*$ threshold.  
This implies there is likely no exotic hadron with these quantum numbers in this energy region, at least at the employed larger-than-physical light quark masses. 

The $DD^*$ scattering amplitude with $I=1$ has been extracted from finite-volume lattice energy levels based on the plane wave approach combined with  the effective field theory. This   incorporates the  one-pion exchange  for the lattice kinematics with $m_\pi>m_{D^*}-m_D$. As expected, the effect of the one-pion exchange is less significant for  the channel with $I=1$ than for the previously explored channel with $I=0$. 

The separate contributions of four Wick contractions to $DD^*$ correlation functions reveal that the contraction denoted by ${\cal C}_2$ in Fig.~\ref{fig:wick} 
plays a key role in distinguishing the isospin channels. Specifically,  it accounts for the repulsion in $I=1$ and attraction in $I=0$. 
This suggests that contributions to $DD^*$ scattering from light-meson exchanges with isovector-vector quantum numbers (such as the  $\rho$-meson exchange) could be significant for the doubly-charmed tetraquark channel. 
 In the framework of chiral EFT, this indicates the potentially important role of isovector-vector  two-pion exchange contributions.

Although it is unlikely to detect bound states or virtual states in the \(I=1\) $DD^*$ channel, as observed in the \(I=0\) channel, the \(I=1\) system still warrant further refined lattice studies alongside the \(I=0\) system. Comparative analyses of $DD^*$ systems with different isospins could provide significant insights into the nature of the interaction of open-flavor, double-heavy systems. Future lattice studies could incorporate analogous simulations across a range of light-quark masses and lattice spacings. As the pion mass approaches its physical value, the \(D^*\) meson becomes unstable and decays via \(D^* \to D\pi\). Consequently, the \(DD\pi\) channel with \(I=1\) would greatly benefit from analyses that explicitly account for three-hadron unitarity, following methodologies similar to those recently applied to the \(I=0\) channel in Refs.~\cite{Hansen:2024ffk,Dawid:2024dgy,Abolnikov:2024key}. Additionally, the ensembles used in this study are limited by relatively coarse lattice spacings, and there is evidence that \(T_{cc}\) solutions may be sensitive to such spacings~\cite{greenTccTetraquarkContinuum2023}. To make definitive conclusions, it will be necessary to repeat these studies using few finer lattice spacings. These considerations underscore the importance of systematically exploring lattice artifacts and ensuring robust results.

\section{Acknowledgements}

We would like to especially thank Sara Collins for producing the charm perambulators and computing the correlators for highest three charm quark masses studied in this work, and for support related to the computer resources used in this project.
We would also like to thank Sara Collins, Alexey Nefediev and Mitja Sadl for discussions. We thank our colleagues in CLS for the joint effort in the generation of the gauge field ensembles which form a basis for the computation. We use the multigrid solver of
Refs.~\cite{Heybrock:2014iga,Heybrock:2015kpy,Richtmann:2016kcq,Georg:2017diz} for the inversion of the Dirac operator. Our code implementing distillation is written within the framework of the Chroma software package~\cite{Edwards:2004sx}. The simulations were performed on the Regensburg Athene2 cluster.   We also thank the HPC RIVR consortium (www.hpc-rivr.si) and EuroHPC JU (eurohpc-ju.europa.eu) for funding this research by providing computing resources of the HPC system Vega at the Institute of Information Science (www.izum.si) and the Department of Theoretical Physics at JSI for the use of the HPC Vega system and the ATOS cluster. We acknowledge support by Slovenian Research Agency ARIS for funding programme P1-0035 and  project J1-3034, by ERC NuclearTheory (grant No. 885150), by BMBF (Contract No. 05P21PCFP1) and by the MKW NRW under the funding code NW21-024-A. We also acknowledge the use of computing clusters at IMSc Chennai. We thank the authors of Ref.~\cite{Morningstar:2017NPB} for making the \textit{TwoHadronsInBox} package public. M.P. gratefully acknowledges support from the Department of Science and Technology, India, SERB Start-up Research Grant No. SRG/2023/001235 and Department of Atomic Energy, India.

\appendix
\section{Interpolators and overlaps}\label{app:interpolators}

 \begin{figure}
	\centering 
		\includegraphics[height=6.cm,width=6.65cm]{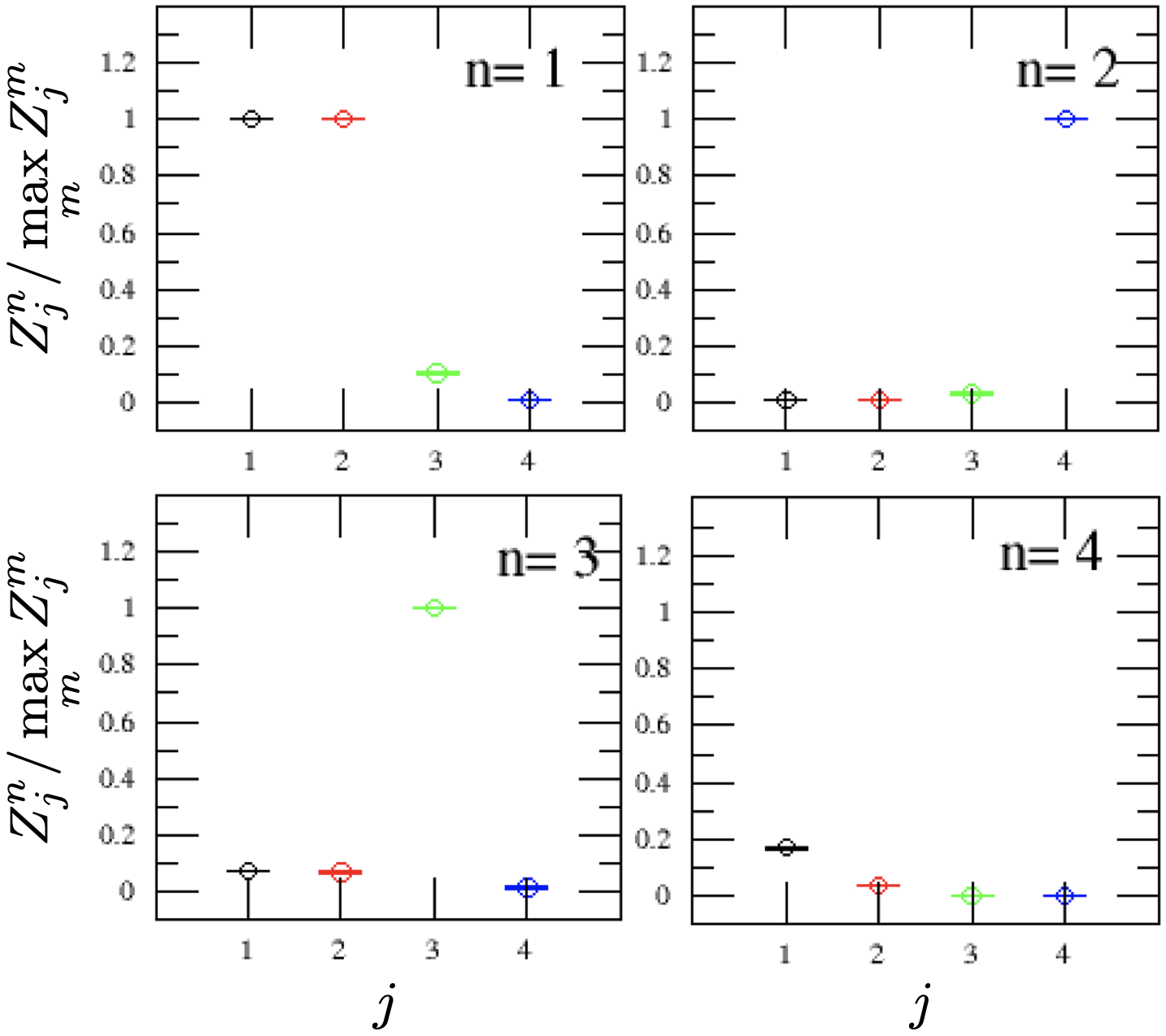}
	 \caption{
 Overlap ratios $ Z_j^n/\max_m Z_j^m $ for the isovector tetraquark channel and lattice extent $N_L=32$. The horizontal axis $j=1,\ldots,4$ represents four operators for irreducible representation  $T_1^+$ with  total momentum zero in Table~\ref{tab:operators}. The ratios are normalized so that they don't depend on the normalization of operators. }  
 \label{fig:overlaps}
\end{figure}

Similarly to Ref.~\cite{Padmanath:2022PRL},
meson-meson interpolators are utilized in this study such that each meson is projected onto a definite momentum

\begin{eqnarray}
 \label{interpolator-I1}
O^{DD^*}_{I=1}
&=&\sum\limits_{k,j}A_{k,j}[D(\vec{p}_{1k})D_j^*(\vec{p}_{2k})]_{I=1}
\nonumber\\
&=& \sum\limits_{k,j}
A_{k,j}[(\bar u\Gamma_{1}c)~(\vec{p}_{1k})~\hspace{0.0cm} (\bar  d \Gamma_{2j}c)~(\vec{p}_{2k})] + \{ u\leftrightarrow d\},  \nonumber\\
\end{eqnarray}
with the appropriate combinations of the Dirac structure  $(\Gamma_{1}, \Gamma_{2j})=(\gamma_5,\gamma_j)$ and  $(\gamma_5\gamma_t,\gamma_j\gamma_t)$. 
Here $\vec{p}_{1k}=p_{1k}\hat{e}_{k}$ and $A_{k,j}$ are overall coefficients from each interpolator. Unlike the isoscalar $T^+_{cc}$ channel, the isovector channel has a plus sign in the second term of Eq.~(\ref{interpolator-I1}). The list with all the employed isovector interpolators  are collected in Table~\ref{tab:operators}.  The flavor symmetry  implies that  the interpolator $[D^*(0)D^*(0)]_{I=1}$  is not present in $T^+_1$ irrep. 

Each eigenstate dominantly couples to a  distinct operator type $D(|\vec p_1^{~2}|)D^*(|\vec p_2^{~2}|)$  as shown by the overlaps $Z_i^{(n)}=\langle O_i|n\rangle$ in Fig.~\ref{fig:overlaps}. 

\begin{table*}[t!]
\centering
\caption{List of the $DD^*$ interpolators for studying the isospin-1  channel at rest and moving frame irreps. In order to classify the operators, here we adopted the notation $O^{D(|\vec{p}_1|^2)D^*(|\vec{p}_2|^2)}_{I=1}$.}
\label{tab:operators}
\begin{ruledtabular}
\begin{tabular}{rc}  
$O^{D(|\vec{p}_1|^2)D^*(|\vec{p}_2|^2)}_{I=1}$ &  \\
\noalign{\smallskip}
\hline
 &$T_1^+$, $\vec{P}= (0,0,0), \hbox{row~z}$  \\
\noalign{\smallskip}
\hline
\noalign{\smallskip}
$O^{D(0)D^*(0)}_{I=1} = $ & $ \bar u\gamma_{5}c~ (\vec{0})~ \bar  d \gamma_{z}c ~(\vec{0}) + \{ u\leftrightarrow d\} $    \\
\noalign{\smallskip}
$O^{D(0)D^*(0)}_{I=1} = $ & $ \bar u\gamma_{5}\gamma_{t}c~ (\vec{0})~ \bar  d \gamma_{z}\gamma_{t}c ~(\vec{0}) + \{ u\leftrightarrow d\} $  \\
\noalign{\smallskip}
$O^{D(1)D^*(1)}_{I=1}|_{\ell=0} = $ & $ \frac{1}{\sqrt{6}} \left[ \bar u\gamma_{5}c~ (\hat{e}_x)~ \bar  d \gamma_{z}c ~(-\hat{e}_x) + \bar u\gamma_{5}c~ (-\hat{e}_x)~ \bar  d \gamma_{z}c ~(\hat{e}_x) \right.$ \\
& $ \left. + \bar u\gamma_{5}c~ (\hat{e}_y)~ \bar  d \gamma_{z}c ~(-\hat{e}_y) + \bar u\gamma_{5}c~ (-\hat{e}_y)~ \bar  d \gamma_{z}c ~(\hat{e}_y) \right.$ \\
& $ \left.+ \bar u\gamma_{5}c~ (\hat{e}_z)~ \bar  d \gamma_{z}c ~(-\hat{e}_z) + \bar u\gamma_{5}c~ (-\hat{e}_z)~ \bar  d \gamma_{z}c ~(\hat{e}_z) \right] + \{ u\leftrightarrow d\}  $  \\ 
\noalign{\smallskip}
$O^{D(1)D^*(1)}_{I=1}|_{\ell=2} = $ & $ \frac{1}{\sqrt{12}} \left[ \bar u\gamma_{5}c~ (\hat{e}_x)~ \bar  d \gamma_{z}c ~(-\hat{e}_x) + \bar u\gamma_{5}c~ (-\hat{e}_x)~ \bar  d \gamma_{z}c ~(\hat{e}_x) \right.$ \\
& $ \left. + \bar u\gamma_{5}c~ (\hat{e}_y)~ \bar  d \gamma_{z}c ~(-\hat{e}_y) + \bar u\gamma_{5}c~ (-\hat{e}_y)~ \bar  d \gamma_{z}c ~(\hat{e}_y) \right.$ \\
& $ \left.-2 \bar u\gamma_{5}c~ (\hat{e}_z)~ \bar  d \gamma_{z}c ~(-\hat{e}_z) -2 \bar u\gamma_{5}c~ (-\hat{e}_z)~ \bar  d \gamma_{z}c ~(\hat{e}_z) \right] + \{ u\leftrightarrow d\}  $  \\ 
\noalign{\smallskip}
\hline
\noalign{\smallskip}
&$A_1^-, \vec P=(0,0,0)$  \\
\noalign{\smallskip}
\hline
\noalign{\smallskip}
$O^{D(1) D^*(1)}_{I=1} = $ & $  \frac{1}{\sqrt{6}} \left[ \bar u\gamma_{5}c~ (\hat{e}_x)~ \bar  d \gamma_{x}c ~(-\hat{e}_x) - \bar u\gamma_{5}c~ (-\hat{e}_x)~ \bar  d \gamma_{x}c ~(\hat{e}_x) \right. $ \\
& $\left. + \bar u\gamma_{5}c~ (\hat{e}_y)~ \bar  d \gamma_{y}c ~(-\hat{e}_y) - \bar u\gamma_{5}c~ (-\hat{e}_y)~ \bar  d \gamma_{y}c ~(\hat{e}_y) \right.$ \\
& $\left. + \bar u\gamma_{5}c~ (\hat{e}_z)~ \bar  d \gamma_{z}c ~(-\hat{e}_z) - \bar u\gamma_{5}c~ (-\hat{e}_z)~ \bar  d \gamma_{z}c ~(\hat{e}_z)\right]  + \{ u\leftrightarrow d\} $  \\ 
\noalign{\smallskip}
$O^{D(1) D^*(1)}_{I=1} = $ & $  \frac{1}{\sqrt{6}} \left[ \bar u\gamma_{5}\gamma_{t}c~ (\hat{e}_x)~ \bar  d \gamma_{x}\gamma_{t}c ~(-\hat{e}_x) - \bar u\gamma_{5}\gamma_{t}c~ (-\hat{e}_x)~ \bar  d \gamma_{x}\gamma_{t}c ~(\hat{e}_x) \right.$ \\
& $ \left.+ \bar u\gamma_{5}\gamma_{t}c~ (\hat{e}_y)~ \bar  d \gamma_{y}\gamma_{t}c ~(-\hat{e}_y) - \bar u\gamma_{5}\gamma_{t}c~ (-\hat{e}_y)~ \bar  d \gamma_{y}\gamma_{t}c ~(\hat{e}_y) \right.$ \\
& $\left. + \bar u\gamma_{5}\gamma_{t}c~ (\hat{e}_z)~ \bar  d \gamma_{z}\gamma_{t}c ~(-\hat{e}_z) - \bar u\gamma_{5}\gamma_{t}c~ (-\hat{e}_z)~ \bar  d \gamma_{z}\gamma_{t}c ~(\hat{e}_z) \right] + \{ u\leftrightarrow d\}  $  \\ 
\noalign{\smallskip}
\hline
\noalign{\smallskip}
&$A_2, \vec P=(0,0,1)$\\
\noalign{\smallskip}
\hline
\noalign{\smallskip}
$O^{D(0) D^*(1)}_{I=1} = $ & $ \bar u\gamma_{5}c~ (\vec{0})~ \bar  d \gamma_{z}c ~(\hat{e}_z) + \{ u\leftrightarrow d\} $    \\
\noalign{\smallskip}
$O^{D(0) D^*(1)}_{I=1} = $ & $ \bar u\gamma_{5}\gamma_{t}c~ (\vec{0})~ \bar  d \gamma_{z}\gamma_{t}c ~(\hat{e}_z) + \{ u\leftrightarrow d\} $  \\
\noalign{\smallskip}
$O^{D(1) D^*(0)}_{I=1} = $ & $ \bar u\gamma_{5}c~ (\hat{e}_z)~ \bar  d \gamma_{z}c ~(\vec{0}) + \{ u\leftrightarrow d\} $    \\
\noalign{\smallskip}
$O^{D(1) D^*(0)}_{I=1}= $ & $ \bar u\gamma_{5}\gamma_{t}c~ (\hat{e}_z)~ \bar  d \gamma_{z}\gamma_{t}c ~(\vec{0}) + \{ u\leftrightarrow d\} $  \\
\noalign{\smallskip}
\end{tabular}
\end{ruledtabular}
\end{table*}

\section{Effective range parameters}\label{app:ere}
The Table~\ref{tab:ERE-observables} provides effective range parameters $a_0$ and $r_0$ for two types of fits shown in Fig.~\ref{fig:fitERE} that were performed in Sec.~\ref{sec:LQC_ERE}.

\begin{table*}[!ht]
 	\caption{Effective range parameters for $DD^*$ scattering from approach in Sec.~\ref{sec:LQC_ERE} using fits $p\cot\delta_0=1/a_0+r_0 p^2/2$ (rows 2-4) and $p\cot\delta_0=1/a_0$ (rows 5-6), which do not incorporate left-hand cut from one-pion exchange.\label{tab:ERE-observables}}
\begin{ruledtabular}
\begin{tabular}{c|c c c c c }  
    Sets & 1  & 2  & 3  & 4  & 5   \\
\hline 
$\chi^{2}/\mathrm{dof}$  & $2.2/6$  & $1.6/6$  & $1.8/6$  & $2.0/6$  & $2.4/6$   \\
$a_0$~[fm] & -0.39($^{+2}_{-3}$) & -0.37($^{+4}_{-4}$) & -0.37($^{+2}_{-1}$) & -0.38($^{+5}_{-4}$) & -0.40($^{+5}_{-3}$) \\
$r_0$~[fm] & -1.13($^{+16}_{-23}$) & -0.62($^{+28}_{-25}$) & -0.61($^{+14}_{-12}$) & -0.61($^{+30}_{-27}$) & -0.60($^{+36}_{-15}$) \\
\hline 
$\chi^{2}/\mathrm{dof}$  & $17.3/7$  & $6.6/7$  & $7.0/7$  & $6.0/7$  & $5.5/7$  \\
$a_0~$[fm]  & $-0.27(^{+7}_{-1})$  & $-0.31(^{+3}_{-3})$  & $-0.33(^{+4}_{-1})$  & $-0.31(^{+3}_{-1})$  & $-0.33(^{+8}_{-2})$ \\ 
\hline   
\end{tabular}
\end{ruledtabular}
\end{table*}

\section{Flavor factors }\label{sec:isospin-factors}

 The isospin factor 
 in Eq.~(\ref{isospins}) related to the contraction ${\cal C}_2$ in Fig.~\ref{fig:wick}b is given by    
\begin{align} 
&|(DD^*)_I\rangle=  \tfrac{1}{\sqrt{2}} | D^0 D^{*+}+ (-1)^{I+1} D^+ D^{*0}\rangle \  ,  \\
&\langle (DD^*)_I | 
{\bm{\tau}^{(1)} \cdot \bm{\tau}^{(2)}}
| (DD^*)_I \rangle_{{\cal C}_2} =\nonumber  \\
&=2 \langle (DD^*)_I | \tau^{(1)} _+\tau^{(2)}_-+\tau^{(1)}_-\tau^{(2)}_+ | (DD^*)_I \rangle \nonumber\\
&= (-1)^{I+1} (   
\langle D^{*0}|\tau_-|D^{*+}\rangle 
\langle D^{+}|\tau_+|D^{0}\rangle + \nonumber\\
&\qquad\qquad  \quad \langle D^{0}|\tau_-|D^{+}\rangle 
\langle D^{*+}|\tau_+|D^{*0}\rangle)\nonumber\\
&=2 \cdot (-1)^{I+1}~.\nonumber
\end{align}
The term $\tau_3^{(1)}\tau_3^{(2)}$ does not contribute to the contraction ${\cal C}_2$ due to the charge conservation in each vertex. 
 We   employed $
 \bm{\tau}^{(1)} \cdot \bm{\tau}^{(2)}
 =\tau_3^{(1)}\tau_3^{(2)} + 2( \tau_+^{(1)}\tau_-^{(2)}+\tau^{(1)}_-\tau^{(2)}_+)$ and   the isospin doublet  $D^{(*)}=(I_3\!=\!\tfrac{1}{2},I_3\!=\!-\tfrac{1}{2})=(-c\bar d, c\bar u) $, while the minus in $-c\bar d$ does not play a role as it appears in bra and ket.

\bibliographystyle{apsrev4-1}
\bibliography{references.bib}
\end{document}